\documentclass[twocolumn]{aastex62}

\usepackage{microtype}
\usepackage{rotating}
\usepackage{booktabs}
\usepackage{threeparttable}
\usepackage{tabularx}
\usepackage{lineno}

\newcommand\com[1]{\textcolor{black}{\textnormal{#1}}}
\graphicspath{{./}{figures/}}

\received{TBD}
\revised{TBD}
\accepted{TBD}
\submitjournal{\textbf{AJ}}

\shorttitle{Asteroid GP}
\shortauthors{Lindberg et al.}

\begin{document}

\title{Characterizing Sparse Asteroid Light Curves with Gaussian Processes}

\author[0000-0003-0588-7360]{Christina Willecke Lindberg}
\affiliation{JHU, 3400 North Charles St., 473 Bloomberg Center for Physics and Astronomy, Baltimore, MD, 21218}

\author{Daniela Huppenkothen}
\affiliation{SRON Netherlands Institute for Space Research, Niels Bohrweg 4, 2333 CA Leiden, The Netherlands}

\author{R. Lynne Jones}
\affiliation{Aerotek, Suite 150, 4321 Still Creek Drive, Burnaby, BC V5C6S7, Canada}
\affiliation{Rubin Observatory, 933 N Cherry Ave, Tucson, AZ 85719, USA,} 

\author[0000-0002-4950-6323]{Bryce T. Bolin}
\affiliation{Division of Physics, Mathematics and Astronomy, California Institute of Technology, Pasadena, CA 91125, USA}
\affiliation{IPAC, Mail Code 100-22, Caltech, 1200 E. California Blvd., Pasadena, CA 91125, USA}

\author{Mario Juri\'{c}}
\affiliation{Department of Astronomy, University of Washington, 3910 15th Avenue NE,  Seattle, WA 98195, USA}
\affiliation{DIRAC Institute, Department of Astronomy, University of Washington, 3910 15th Avenue NE, Seattle, WA 98195, USA}

\author[0000-0001-8205-2506]{V. Zach Golkhou}
\affiliation{DIRAC Institute, Department of Astronomy, University of Washington, 3910 15th Avenue NE, Seattle, WA 98195, USA} 

\author[0000-0001-8018-5348]{Eric C. Bellm}
\affiliation{DIRAC Institute, Department of Astronomy, University of Washington, 3910 15th Avenue NE, Seattle, WA 98195, USA}

\author{Andrew J. Drake}
\affiliation{Division of Physics, Mathematics and Astronomy, California Institute of Technology, Pasadena, CA 91125, USA}

\author{Matthew J. Graham}
\affiliation{California Institute of Technology, 1200 E. California Blvd, Pasadena, CA 91125, USA}

\author[0000-0003-2451-5482]{Russ R. Laher}
\affiliation{IPAC, California Institute of Technology, 1200 E. California Blvd, Pasadena, CA 91125, USA}

\author[0000-0003-2242-0244]{\com{Ashish~A.~Mahabal}}
\affiliation{Division of Physics, Mathematics and Astronomy, California
Institute of Technology, Pasadena, CA 91125, USA}
\affiliation{Center for Data Driven Discovery, California Institute of
Technology, Pasadena, CA 91125, USA}

\author[0000-0002-8532-9395]{Frank J. Masci}
\affiliation{IPAC, California Institute of Technology, 1200 E. California Blvd, Pasadena, CA 91125, USA}

\author[0000-0002-0387-370X]{Reed Riddle}
\affiliation{California Institute of Technology, 1200 E. California Blvd, Pasadena, CA 91125, USA}

\author[0000-0002-1486-3582]{Kyung Min Shin}
\affiliation{California Institute of Technology, 1200 E. California Blvd, Pasadena, CA 91125, USA}




\begin{abstract}

In the era of wide-field surveys like the Zwicky Transient Facility and the Rubin Observatory's Legacy Survey of Space and Time, sparse photometric measurements constitute an increasing percentage of asteroid observations, particularly for asteroids newly discovered in these large surveys. Follow-up observations to supplement these sparse data may be prohibitively expensive in many cases, so to overcome these sampling limitations, we introduce a flexible \com{model based on Gaussian Processes to enable Bayesian parameter inference of asteroid time series data}\footnote{code can be found at \href{https://github.com/dirac-institute/asterogap}{AsteroGaP}}. \com{This model is designed to be flexible and extensible, and can model multiple asteroid properties such as the rotation period, light curve amplitude, changing pulse profile, and magnitude changes due to the phase angle evolution at the same time. Here, we focus on the inference of rotation periods.} Based on both simulated light curves and real observations from the Zwicky Transient Facility, we show that the new model reliably infers rotational periods from sparsely sampled light curves, and generally provides well-constrained posterior probability densities for the model parameters. \com{We propose this framework as an intermediate method between fast, but very limited period detection algorithms and much more comprehensive, but computationally expensive shape modeling based on ray-tracing codes.}

\end{abstract}

\keywords{Gaussian Processes --- 
asteroids -- sparse data sets -- etc.}


\section{Introduction} \label{sec:intro}

Asteroids are small rocky bodies known to contain information about the formation of planetary objects in our Solar System. Their shapes, sizes, spin, orbits, and compositions all help us constrain models for asteroid formation, collision history, and orbital evolution. These models influence our understanding of the proto-planetary disk along with the evolution and distribution of materials such as water in our Solar System.

Asteroids are observed in the light they reflect from the sun. The amount of light observed by a telescope on Earth at a given point in time depends on the illuminated asteroid surface visible by the telescope as well as surface properties such as composition and albedo. Because asteroids move with respect to both Earth and Sun, and also rotate around their own axis, the observed flux will change as a function of time \citep{barucci1982}. 

Consequently, asteroid light curves can be described by parameters like rotational period, amplitude of the magnitude variations, and the variability in magnitude within a rotational period (which we denote \com{as} \textit{rotational profile} in this paper). Determining these parameter values can constrain physical asteroid characteristics such as elongation, shape, and structural properties (e.g.~ whether it is a rubble pile vs.~a monolithic rock, \citet{Masiero2009}). 

A standard ellipsoidal asteroid will produce a light curve that is roughly sinusoidal and double-peaked, which can easily be modeled with an ensemble of sine functions and characterized by a Fourier transformation \citep{Harris2014}. However, if an asteroid is not strictly tri-axial or is composed of surface materials with varying albedos, then rotational profile will become complex. While one can still derive Fourier representations of these complex light curves, their interpretation is much more difficult, since the variability within each period will spread power across a wide range of frequencies. If the data is sparse, it then becomes difficult to derive an accurate rotational period. Recently, \citet{schemel2020} introduced a method to infer certain asteroid characteristics (e.g. color, phase, parameters, absolute magnitude, amplitude of rotation) with sparse photometric data by assuming that the light curve can be described by a single sinusoid, however, this leaves numerous other asteroid characteristics unconstrained. 

Asteroid research has for the most part utilized phase dispersion minimization \citep{phase_dispersion_min} or Fourier transform-based methods like Fourier Analysis of Light Curves (FALC) \citep{Harris1989} and Lomb-Scargle periodograms \citep{lsp_lomb, lsp_scargle} to characterize rotational periods and amplitudes. It is not uncommon that groups relying on differing densities of observations will report significantly different period estimates for the same object when relying on Fourier-based methods \citep{Fedorets2020, Bolin2020}. These approaches have been shown to be less reliable when light curves are sparsely sampled, producing multi-peaked periodograms \citep{Masiero2009, Warner2011, Harris2012} and resulting in period estimates that can differ from the period derived through other data sets or methods. One potential solution is to gather denser time series data, which is expensive, and may be infeasible for large samples of asteroids. 

Future asteroid time series data will primarily be collected from large surveys with telescopes such as the Rubin Observatory, which is expected to observe 500,000 asteroids per night with two observations per asteroid \citep{LSST_asteroids, Ivezic2019}. Likewise, the Zwicky Transit Facility (ZTF) detects ~25,000 asteroids on a clear night, covering the northern sky every three nights \citep{bellm2019, ztf_per_night}.  Gathering dense light curves is an impractical task for these large new samples of asteroids. 
Hence, reliable inference of the properties of the millions of asteroids to be detected will require adequate methods to characterize these sparse light curves. Here, we propose a new model based on Gaussian Processes to overcome these obstacles and provide an alternative approach to characterizing asteroid light curves.

Gaussian Processes \citep{rasmussen} provide flexible and generative models for time series data, and have been used to model, among others, non-sinusoidal light curves of stars \citep{brewer&stello} and Active Galactic Nuclei \citep{kelly2014}. This is done by modeling the covariance between data points, rather than modeling the light curve directly. 
In this paper, we show that we can obtain accurate estimates of asteroid properties with the use of a Gaussian Process model and sampling via Markov Chain Monte Carlo (MCMC). \com{Gaussian Processes have a number of favourable properties when considering models of asteroid light curves. (1) They permit periodic, non-sinusoidal functions that can be used to encode the rapid variability seen from many asteroids. (2) They allow straightforward modeling of changes in the rotational profile over time as the viewing angle between asteroid and observer changes, as well as phase angle modulations through appropriate modeling terms. (3) They enable simultaneous inference of other asteroid properties (e.g. the light curve amplitude and intra-period variability) along with the period. (4) They are heavily used in forecasting, and can provide probabilistic forecasts for when to schedule the most informative follow-up observations.}

The development of this model was initially motivated by sparse observations of the interstellar object 1I/'Oumuamua, observed in late 2017 \citep{Bolin2018}. Using photometric time-series data from the Apache Point Observatory (APO) and the Discovery Channel Telescope (DCT), \citet{Bolin2018} constructed a light curve and estimated a rotational period. In this context, a Gaussian Process yielded a well-constrained estimate of the period despite the sparsity of the data. It is with this sparsity in mind that we continue to develop and validate this Gaussian Process method. \com{In this paper, we focus primarily on generating probabilistic estimates of the asteroid rotational period as it is both a key asteroid characteristic as well as a quantity for which numerous standard methods are available for comparison. Where appropriate, we also highlight where the generative nature of the Gaussian Process can be used e.g.~for population studies, and how other parameters might be exploited to explore other asteroid properties such as its shape.} 

The structure of this paper is as follows: In Section \ref{sec:methods}, we briefly review the concept of Gaussian Processes, covariance functions, Bayesian priors, and MCMC samplers used in our model. In Section \ref{sec:sim}, we outline our simulations and give a motivating example of our model being used on a well-known nearby asteroid. In Section \ref{sec:ztf}, we show results for asteroids recently observed with ZTF, and in Section \ref{sec:discus}, we discuss the potential of this model and describe future improvements.


\section{Methods} \label{sec:methods}

\subsection{Bayesian Statistics} \label{sec:bayes}


In Bayesian inference, we estimate a probability distribution of model parameters, given observed data and prior knowledge about the model and its parameters.
Bayes' theorem is defined as
\begin{equation}
    p(\theta|D, M) = \frac{p(D|\theta, M)p(\theta | M)}{p(D | M)} \; .
\end{equation}
 
\noindent In this equation, $p(\theta | D, M)$ denotes the \textit{posterior probability distribution} of the model parameters $\theta$, given an observed set of data points $D$ and an assumed model $M$. This assumed model $M$ deviates from standard astronomical terminology in that its definition includes not only a physically motivated function to model the data, but also a number of implicit assumptions, including the type of equation used to represent the data points--in our case a Gaussian Process--as well as the model for the uncertainty on the data points (the likelihood) and other choices made during the data analysis process. The right-hand side of the equation defines two important components for inference of the parameters $\theta$. The \textit{likelihood} describes the probability of observing a data set $D$, given some underlying model $M$ with parameters $\theta$, which are assumed to be known exactly. In practice, we generally observe data and aim to infer the parameters, which makes the likelihood a \textit{function} of the parameters, but a \textit{probability distribution} of the data. The likelihood includes both a parametric model of the underlying (physical) process assumed to have generated the data, as well as a model for the measurement uncertainties. 

The prior probability distribution, $p(\theta | M)$ encodes our knowledge of the parameters $\theta$ before looking at the data. For example, asteroids cannot have arbitrary rotation periods. The possible values for the rotational period of an asteroid are constrained both by physics--for example the material the asteroid is made of--and prior knowledge about the range of periods we have observed from similar asteroids in the past. We encode that information in the form of prior probability distributions, specified in detail in Section \ref{sec:priors}. 

The final term in Bayes' theorem in the denominator is called the \textit{evidence} or \textit{marginal likelihood}. It is a normalization term ensuring that the posterior $p(\theta | D, M)$ integrates to unity as a proper probability distribution. While important for the purposes of model selection, it is independent of the parameters $\theta$, and is a constant that that does not affect the shape of the posterior, only its normalization. We can therefore write Bayes' theorem as 

\begin{equation}
    \label{eqn:postprop}
     p(\theta|D, M) \propto p(D|\theta, M)p(\theta | M)
\end{equation}

\noindent when performing inference on the parameters.

In most practical applications, the posterior defined in Equation \ref{eqn:postprop} is analytically intractable and needs to be approximated numerically. One way to do this is through Markov Chain Monte Carlo (MCMC) sampling. 
In the ideal case of an infinitely long sampling process, MCMC guarantees convergence to the correct posterior distribution. During an initial burn-in phase, an MCMC chain--a single sampling process--should converge to a stationary distribution and ideally, any samples subsequently generated are drawn from the true posterior distribution. Since infinite MCMC runs are impossible, we run the sampler for a specified number of iterations, and check for convergence---using autocorrelation times and the visual inspection of MCMC chains---to see whether the chains have converged to the posterior distribution. For sparse asteroid observations, we expect the posterior probability distribution for the period to be multimodal, i.e.~exhibit multiple peaks. This is caused by the sparsity of the data, which admits multiple periods as valid models. It is therefore difficult for any sampler to sample efficiently since chains may become constrained within their localized probability maxima and never get a chance to sample other regions of the parameter space. In this work, we use MCMC as implemented in the python package \textit{emcee} \citep{emcee}. 






\begin{figure*}
    \centering
    \includegraphics[width=\textwidth]{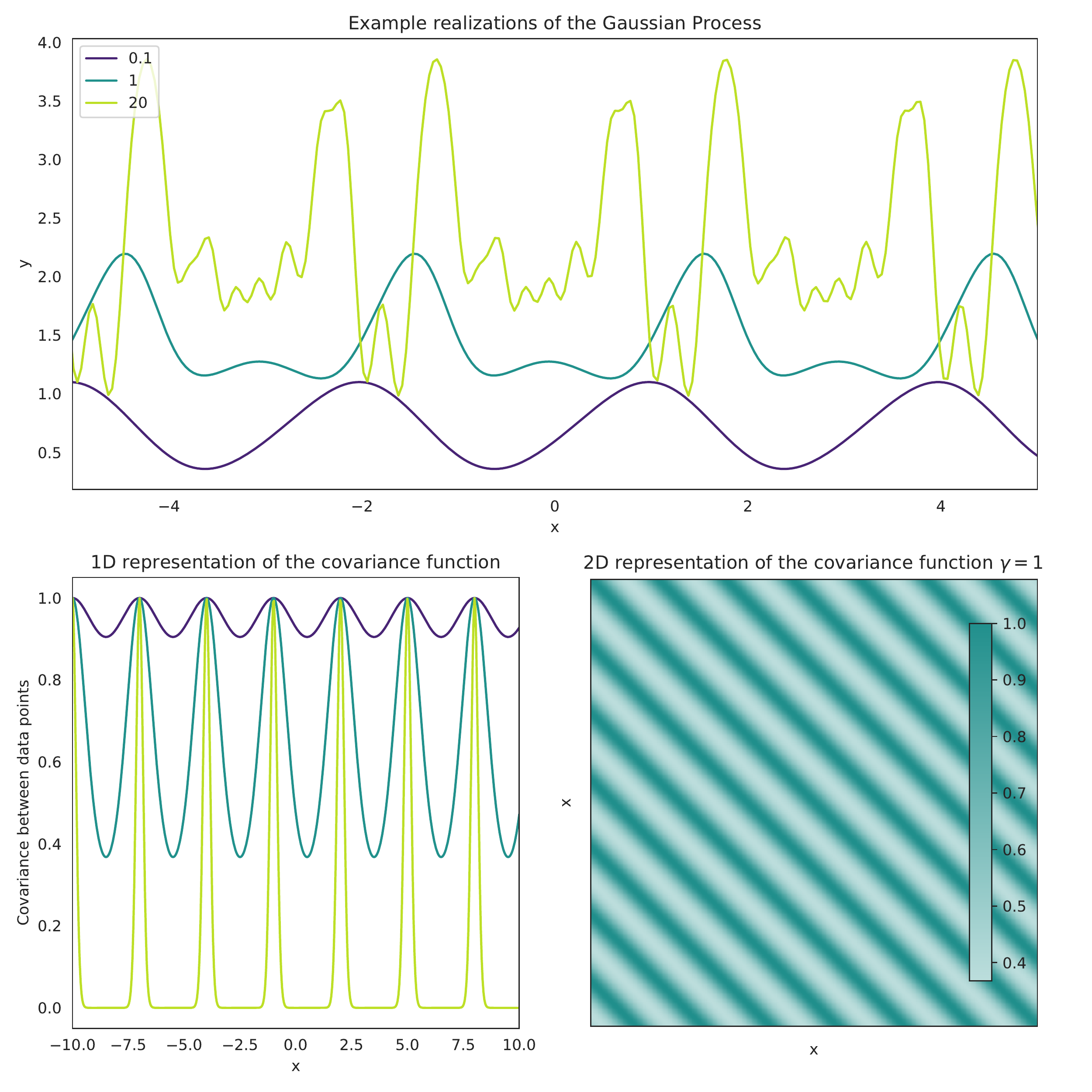}
    \caption{Top: Three example realizations drawn from a Gaussian Process with a periodic covariance function. All three realizations were drawn from a Gaussian Process with a period of  $P=3.0$ and an amplitude of $A=1$, but variable intra-period lengthscale $\Gamma$. Bottom left: visualization of the covariance function that generated the realizations in the top panel. Given the period of $P=3$, points that are integer multiples apart from each other will be exactly correlated, and thus generate a periodic pattern. The variations within each period depend on the value of $\Gamma$: if $\Gamma$ is large, the correlation between points that are close together is near zero, generating rapidly varying structure within each period. If $\Gamma$ is small, the correlation between data points is always high, generating smooth, nearly sinusoidal realizations. Bottom right: two-dimensional representation of the covariance function with $\Gamma=20$ from the left panel. The light yellow bands are regions of high covariance spaced at intervals of multiples of $P$, the dark purple bands in between show regions where the covariance between data points for a given $\Delta x$ is near $0$. }
    \label{fig:covariance}
\end{figure*}
\subsection{Gaussian Processes}\label{sec:gp}

Gaussian Process models are a generative kernel-based framework for solving both classification and regression problems \citep{Roberts2012}; here, we focus on Gaussian Processes as a method for regression. Rather than modeling the data points directly as in standard linear regression, a Gaussian Process provides a model for the \textit{covariance} between data points. 
Let us assume we have obtained $N$ photometric measurements $D = \{ m_i, \sigma_m\}_{i=1}^{N}$ of an asteroid, recording magnitude \com{and} magnitude error for every data point at times $T = \{t_i\}_{i=1}^{N}$. 

In standard linear regression with a Gaussian likelihood\footnote{also called $\chi^2$ statistic in many astronomical contexts}, one defines a parametric model $f(t_i, \phi)$ specified by a set of parameters $\phi$ that aims to approximate the structure in the observed data points. One then compares model magnitudes derived for a specific set of values for $\phi$ to the observations using a Gaussian likelihood, where the mean of the Gaussian is set by $f(t_i, \phi)$, the variance is set by measurement uncertainties, $\sigma_m^2$, and data points are assumed to be independent random variables, i.e. there are no covariances between data points. It is possible to write this model as a single Gaussian distribution, 

\[
D \sim \mathcal{N}(f(T,\phi), \Sigma^2)
\]

where $D$ is the vector of $N$ data points, $\mathcal{N}$ is an $N$-dimensional Gaussian distribution with means set by $f(T, \phi)$ and $\Sigma$ is an $N$x$N$ dimensional matrix where, in this standard case, the off-diagonal elements are zero and the diagonal elements are set by $\sigma_m^2$.

A Gaussian Process extends this model by allowing covariances between data points to exist, and by defining a model for that covariance. It is a generalization of the standard Gaussian likelihood into a process that defines an infinite-dimensional Gaussian distribution. One may still define a mean function $f(t, \phi)$ that governs global trends in the data set, but this model now also contains a model for the covariances, called a covariance function or a kernel, $k(t_i, t_j, \theta)$ with its own set of parameters $\theta$. Instead of drawing a vector as one would from a finite Gaussian distribution, draws from a Gaussian Process are functions defined by the mean function $f$ and covariance function $k$. The draws, evaluated in finite dimensions and conditioned on the observations, serve as model light curves for the observed data points.

Modeling the covariance between every pair of data points through the covariance function allows us to build flexible models in particular for stochastic processes. Here we assume that the small-scale structure and variations in albedo of an asteroid, generating rapid magnitude variations, can be approximated by such a stochastic process. 


The logarithm of the likelihood for the Gaussian Process model is defined as

\begin{eqnarray}
    \ln \mathcal{L}(\phi, \theta) & = & \ln p(D|T,\phi,\theta) \nonumber \\ 
     & = & -\frac{1}{2} r_{\phi} K_{\theta}^{-1} r_{\phi} - \frac{1}{2} \ln \det K_{\theta} - \frac{N}{2} \ln(2\pi) \label{eq:lnL}
\end{eqnarray}
where
\begin{equation}
    r_\theta=(m_1 - f(t_1,\phi) \dots m_N - f(t_N, \phi)^T
\end{equation}

\noindent represents the vector of residuals and $K_\theta$ represents the covariance matrix (\citealt{rasmussen}; see also \citealt{celerite}). The covariance matrix is defined as the $N\times N$-dimensional subset of the infinite-dimensional covariance function $k$ for all observation time stamps $(t_i, t_j)$.


The covariance function defines the covariance matrix $K_\theta$ as a function of parameters $\theta$ and thus describes how pairs of data points are expected to covary under the assumptions of the model. There are both stationary and non-stationary covariance functions. Stationary covariance functions are often defined in terms of the distance between data points: $k(t_i, t_j) = k(|t_i - t_j|)$ such that the statistical properties of the data-generating process do not change as a function of time. Covariance functions constructed with a symmetric function that generates periodic processes are a special case of non-stationary covariance functions. 

In this work, we use the \textit{sine-squared exponential covariance function}, which requires that the covariance between data points spaced at a certain periodic intervals $P$ is unity. This means that for any given point in time $t_i$, data points at multiples of $P$ from $t_i$ will be exactly equal, while data points at a distance of $t_i+\Delta t$, where $\Delta t < P$ may be correlated or uncorrelated. This defines a process that is strictly periodic, but not sinusoidal, allowing for rapid changes within a period. This covariance function is formally defined as

\begin{equation}
k_{\mathrm{per}}(t_i, t_j) = A_\mathrm{periodic}  \exp(-\Gamma \sin^2[\frac{\pi}{P} |t_i-t_j|])
\label{eqn:sinesquaredkernel}
\end{equation}

The behaviour of this covariance function is governed by a number of parameters\footnote{In the machine learning literature, the kernel parameters are often called "hyperparameters", because they kind of consider the model data points as the actual parameters. However, in Bayesian inference, "parameters" generally refers to the parameters of the model, and hyperparameters are the parameters used to define the prior distributions. In this paper, we will consistently use Bayesian inference terminology.}. The amplitude $A_\mathrm{periodic}$ scales the covariance function to match the amplitude of the light curves (see also Table \ref{table:priors}). $\Gamma$ is defined as the inverse of a length scale that dictates the distance between data points within a single period at which correlations tend to zero. This models the intra-period variability of our light curve, which can reveal information about the asteroid shape and albedo variations. Finally, $P$ denotes the period and indicates the interval where the correlation between data points becomes unity. In this work, we use this parameter as a proxy for the rotational period of the asteroid. We expect that the sine-squared exponential covariance function provides an adequate representation for asteroid light curves, since it accounts for intra-periodic variability and assumes our data is periodic without necessarily being sinusoidal. However, we also note that the strict periodicity requirement it imposes may make the model unstable for light curves spanning very long intervals (e.g. years), \com{binary,} or tumbling asteroids that rotate around non-principal axes. 

In Figure \ref{fig:covariance}, we show examples of time series generated by a Gaussian Process with different values of $\Gamma$, along with the corresponding covariance functions and behaviour. For $\Gamma=0.1$, the covariance function is never zero, i.e. the covariance between data points is always high, and data points are never independent. This leads to very smoothly varying functions. In contrast, for larger values of $\Gamma$, the covariance function tends to $0$ in between multiples of $P$, which allows the model to produce more complex, variable functions. 

Asteroids exhibit light curves with significant intra-period variability, which we aim to model with the covariance function presented in Equation \ref{eqn:sinesquaredkernel}. For long observations, the rotational profile changes as a function of time: as the angle between Earth, Sun and asteroid changes, so does the reflecting surface visible from Earth. The covariance function generates strictly periodic function with the same rotational profile throughout. In order to model the nature of a slowly varying rotational profile, we need to add another component to the model.

A covariance function that is often used is the \textit{squared exponential covariance function}, a stationary covariance function that models the covariance as an exponentially decreasing function of distance between data points, i.e., observations taken close together are expected to be more highly correlated than observations taken far apart. This covariance function is formally defined as

\begin{equation}
k_{\mathrm{SE}} = A_\mathrm{long} \exp(-\frac{|t_i-t_j|^2}{2M})
\label{eqn:squaredkernel}
\end{equation}

The metric $M$ determines the length scale of the correlation between data points taken at a distance of $|t_i-t_j|$ from each other. A smaller metric leads to a correlation that drops off faster than a larger metric.

The product of this covariance function with the sine-squared exponential covariance function allows the Gaussian Process to generate functions that change the shape of their intra-period variability over time, governed by the length scale $M$. We use this combination of covariance functions to mimic the natural evolution of an asteroid rotational profile and its phase-angle between the Earth and the Sun changes during the course of its orbit.
 
Given these covariance functions, we can now compute an $N\times N$ covariance matrix $K_\theta$ for each of the $N$ data points in the data set, and generate model light curves conditioned on the observed data. We include the magnitude errors into the model by adding a noise constant to the covariance matrix along the diagonal:

\begin{equation}
    k_{\mathrm{WN}}(t_i, t_j) = \delta_{ij}\sigma_i^2 \; .
\end{equation}

\noindent Our final model includes a mean function $f(t, \phi) = c$ which we set to be constant, and the final covariance function

\begin{equation}
    k(t_i, t_j) = k_\mathrm{SE}(t_i, t_j) k_\mathrm{per}(t_i, t_j) + k_\mathrm{WN}(t_i, t_j) \; .
\end{equation}

\noindent We use this model to estimate the likelihood defined in Equation \ref{eq:lnL}. Here, we use Gaussian Processes implemented in the Python package \textit{george} \citep{george}. 

\begin{table*}[hbtp]
\renewcommand{\arraystretch}{1.3}
\footnotesize
\caption{Overview of the priors used in the Bayesian models}
\begin{threeparttable} 
\begin{tabularx}{18cm}{p{6cm}p{3cm}p{3cm}p{8cm}}
\toprule
\textbf{Parameter Name} &\textbf{Parameter}  & \textbf{Prior distribution}   & \textbf{Hyperparameters}          \\\midrule
Mean constant magnitudes$^\emph{a}$ & $c$      & normal  & $\mu_m = 1, \sigma_m=0.5$                         \\
Periodic kernel, amplitude & $\log{A_\mathrm{periodic}}$   & normal  & $\mu_{A,p} = log(0.15), \sigma_{A,p}=log(2)$              \\
Periodic kernel, inverse length scale & $\log{\Gamma}$  & normal  & $\mu_\Gamma= log(10), \sigma_\Gamma=log(2)$                \\
Periodic kernel, period & $\log{P}$ & kernel density estimate  & bandwidth = 0.2 / $\sigma(\mathrm{data})^\emph{b}$             \\
Squared Exponential kernel, amplitude & $\log{A_\mathrm{long}}$ & uniform  & $[-10, 10 ]$            \\
Squared Exponential kernel, length scale$^\emph{c}$ & $\log{M}$ & normal  & $\mu_M = log(100), \sigma_M=(10)$            \\

\bottomrule
\end{tabularx}
   \begin{tablenotes}
      \item[\emph{a}]{Simulated magnitude data is normalized to 1 and ZTF magnitude data is corrected to 0.}
      \item[\emph{b}]{$\sigma(\mathrm{data})$ here refers to the standard deviation of the data.}
      \item[\emph{c}]{This is often also referred to as \textit{metric}, which we will use below as the parameter name to easily distinguish it from the inverse length scale of the periodic kernel.}

\end{tablenotes}
\end{threeparttable}
\label{table:priors}
\end{table*}

\subsection{Priors}\label{sec:priors}

Using a Gaussian Process with a constant mean function, a squared exponential covariance function, and a sine-squared exponential covariance function, we need to define prior distributions for six parameters, defined in Table \ref{table:priors}.

\begin{figure}
\includegraphics[width=9.5cm]{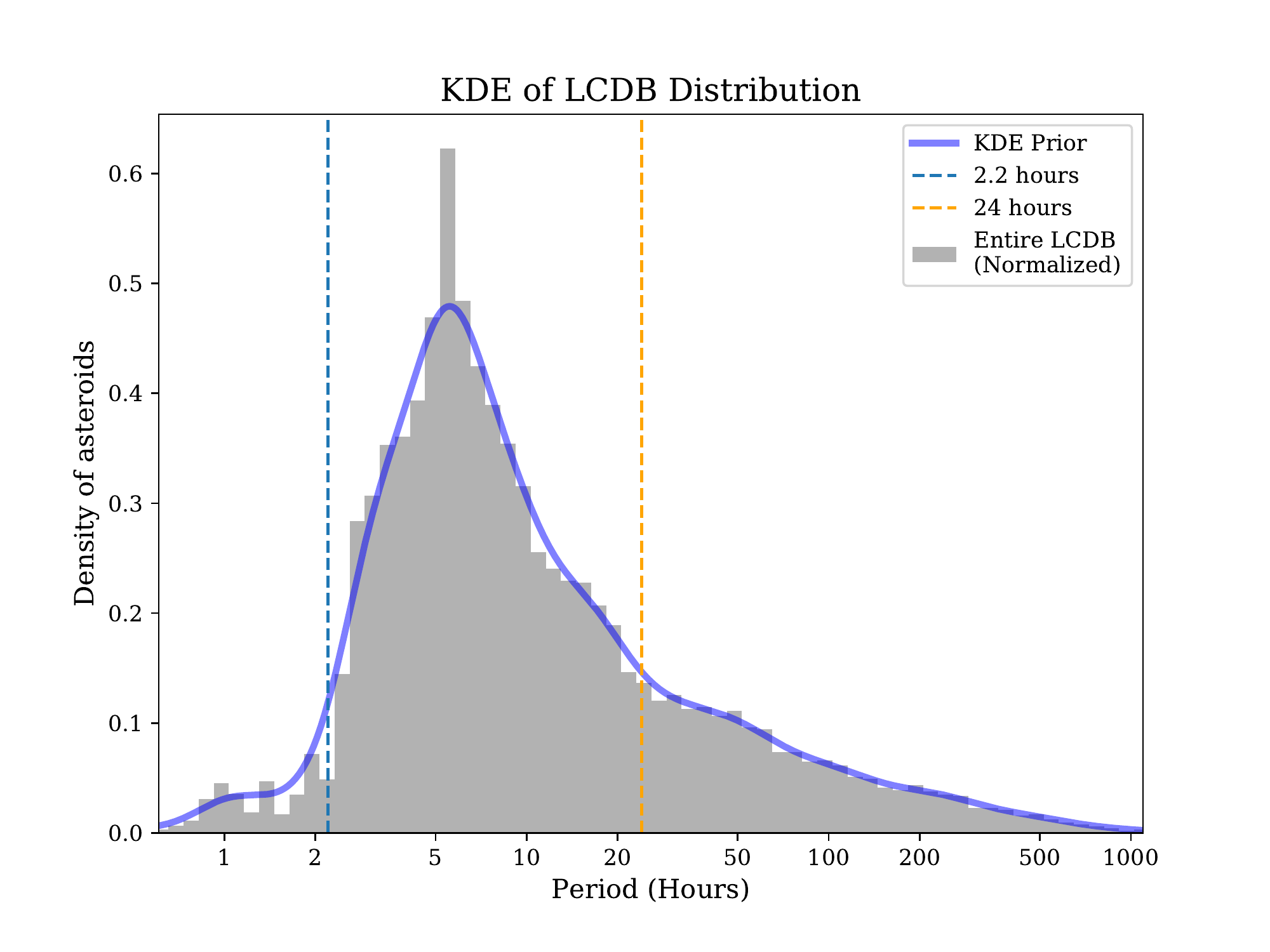}
\centering
\caption{This figure shows the normalized distribution of recorded asteroid period values from the Lightcurve Database (LCDB) as of March 2020 (grey) and the kernel density estimate distribution of our period prior (blue).}
\label{fig:lcdb}
\end{figure}

We generate a prior probability density distribution for the period parameter $P$ by approximating the observed distribution of asteroid rotational periods derived from the sample of 33969 known asteroids available in the Lightcurve Database \citep{LCDB} as of March 25th 2020, shown in Figure \ref{fig:lcdb}. We use a Gaussian kernel density estimate (KDE) \citep{kde1, kde2} as implemented in \texttt{SciPy} \citep{scipy} to approximate this distribution. We implement hard cutoffs in our prior to only allow periods longer than 1 minute and shorter than one year (8760 hours) to avoid unphysical period suggestions. We note that this prior distribution is unlikely to adequately reflect the true distribution of observed periods, but rather a convolution of that distribution with a number of observational biases \citep{Jedicke2016, Masiero2009}. We considered two alternative priors: a uniform distribution and an exponentially decaying distribution. Because the rotational periods of asteroids can span multiple orders of magnitude--from under an hour to hundreds of hours--we discarded the option of a flat (uniform) prior early on, since it would disproportionally favour very long periods, unlike what is seen in practice. Conversely, an exponentially decaying prior (often implemented as a uniform prior on the logarithm of the relevant parameter), would strongly favour extremely short periods. We arrived at the empirically derived distribution above as a compromise: while it might encode some observational biases, it allows for a wide range of possible periods, and correctly assigns a large probability where we expect most asteroid rotational periods to exist. For this work, where we infer the properties of individual asteroids, we expect that this prior will suffice. However, we caution the reader that using this model e.g. for population inference will require a more nuanced consideration of the period prior.  


We impose a strict positivity requirement on both amplitude parameters, $A_\mathrm{long}$ and $A_\mathrm{periodic}$, which by definition must not be negative. We use the logarithms of the amplitudes, $\log(A)$ as parameters rather than $A$ itself, and assign a uniform (for $A_\mathrm{long}$) and a normal distribution (for $A_\mathrm{periodic}$) to these parameters, respectively. Hyperparameter values for both are chosen to reflect magnitude changes broadly typical for asteroid observations. The periodic lightcurve amplitude, $A_\mathrm{periodic}$, may increase or decrease for a given object based on its phase angle \citep{Zappala1990}  or its surface properties and topography \citep{Gutierrez2006}. The dynamic range of $A_\mathrm{long}$ will depend on the class of asteroid and how much their viewing geometry evolves in time. For asteroids closer to the Earth, where their viewing geometry is more drastically and rapidly evolving in a given apparition, the dynamic range is expected to be larger compared to more distant objects like Kuiper Belt objects, which have more consistency in their apparition. 

For science cases other than the case considered here, the choice of hyperparameters can be tuned to fit the physical knowledge of other light curves exhibiting periodicity. The priors with their hyperparameters are summarized in Table \ref{table:priors} and a visualization can be found online in the \href{https://github.com/dirac-institute/asterogap}{AsteroGaP documentation}.

\subsection{Assumptions} \label{sec:assumptions}

The model we describe above implicitly and explicitly makes a number of assumptions about the data we observe \com{and} the inference problem of asteroid periods. Because these assumptions are important to the validity of the model, we list them here. For cases where any of these assumptions are broken, a different model, or the addition of other model components, may be required.

\begin{enumerate}
  \item We assume long-term trends due to the changing distance to the asteroid between the asteroid and the Sun have been removed. Our model does account for gradual changes in the rotational profile from the changes in phase-angle, but the constant mean function $c$ implies that any major changes in the mean magnitude for our asteroids have been normalized out, e.g. by using a software that simulates the asteroid's orbit and distance from the Sun like OpenOrb \citep{openorb}. \com{This assumption can be relaxed by inclusion of a phase angle model in the mean function.}
 

  \item We assume that our asteroid light curve data is strictly a periodic process, meaning we cannot account for tumbling asteroids which exhibit rotation along a non-principal axis, or asteroid observations taken over many years, where the period might change. \com{This assumption can be relaxed by implementing a covariance function that is not strictly periodic. Examples of quasi-periodic covariance functions can be found for example in \citet{celerite}.}
  \item The Gaussian Process, by its nature, requires measurement uncertainties to be Gaussian, which should generally be true for optical magnitudes from astronomical surveys, but might break down for other types of data.
\end{enumerate}

\section{Simulated Data}\label{sec:sim}

We evaluate the effectiveness of our Gaussian Process model on simulated light curves of asteroids, generated from asteroid shape models with established rotational period. With these simulations, we aim to show that the Gaussian Process can recover rotational periods in realistic simulations of sparsely sampled asteroid time series. We compare the Gaussian Process model with the more conventional Lomb-Scargle Periodogram (LSP) and test how effective each of the two methods are at rediscovering the true \com{(simulated)} period in these simulations. 

\subsection{Simulation Set-Up}\label{sec:sim_setup}

We use three-dimensional asteroid shape models available in the DAMIT database\footnote{see \url{http://astro.troja.mff.cuni.cz/projects/damit}} \citep{DAMIT} to generate realistic, dense time series data spanning 5.5 months (165 days ranging from MJD 49627 to 49787) sampled every 30 seconds for two short-period asteroids, 221 Eos and 3200 Phaethon. \com{We assign magnitude uncertainties equal to 10\% of the standard deviation of all the 30-second magnitude measurements within the varying range of days we test.} Both asteroids have been densely observed multiple times since their discovery and have well-established rotation periods and physical models \citep{LCDB}.

\begin{figure*}
    \centering
    \includegraphics[width=\textwidth]{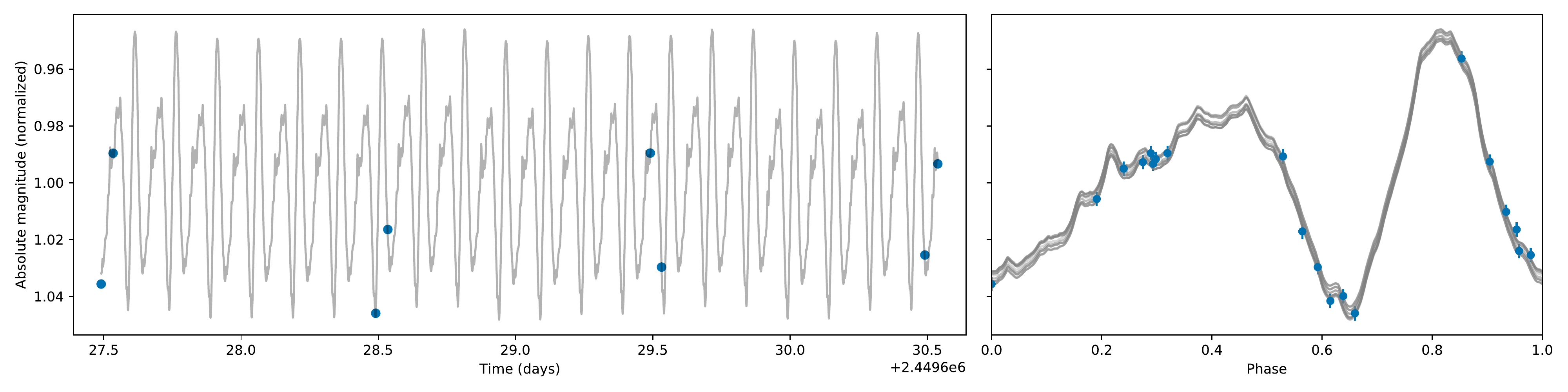}
    \caption{Asteroid 3200 Phaethon light curve with simulated observations over 10 days total. Magnitudes are simulated without phase functions and normalized to 1. Grey lines denote the densely simulated light curve of 3200 Phaethon from the shape model available in the DAMIT database, and the blue markers are the simulated photometric observations obtained from the light curve. Left: 8 observations over the first 4 days of observations, shortened from 10 days for viewing clarity. Right: the light curve for all 10 days of observations, folded at the asteroid's estimated rotational period. }
    \label{fig:3200_unfolded}
\end{figure*}

We sub-sample our dense simulations, creating a cadence pattern similar to ZTF and proposed LSST cadences \citep{Bellm2019bellm, LSST_asteroids}. 

A ZTF and LSST-like cadence consists of 2 observations per night spaced 30-60 minutes apart, every few nights. We simulate two observation every night for $L$ days, such that the number of data points in each light curve $N = 2L$. Intervals between each set of nightly observations were randomly determined by sampling a normal distribution with a mean of 60 minutes and a standard deviation of 5 minutes.  We chose $L$ in a range from 10 to 60 days to test how well our model and the LSP will perform with different numbers of data points. 


We initialize the MCMC sampler with 100 chains for a burn-in of 1,000 iterations, followed by an additional 10,000 iterations to sample. In individual cases, we increase the number of burn-in iterations when the distribution appears to not have converged.
For comparison, we use a two-term multi-band periodic LSP as implemented in the \textit{Python} package \textit{gatspy} \citep{gatspy}. For the short-period asteroids simulated here, we set the period range for the search to span 1 to 60 hours and we set our first pass coverage to 200.

Preliminary experiments reveal posterior distributions that are multimodal in the period parameter, suggesting that summaries of the distribution like the mean and standard deviation fail to capture a complete picture of the posterior inference. In order to identify and characterize modes in the marginalized posterior probability distribution, we initially generate a coarse 20-bin histogram across the full span of sampled parameters \com{and} identify the bins with the highest probability density as potential modes. We produce high-resolution histograms for each of the identified modes, and identify the highest peak in each as the putative mode. We calculate the width to half of the max of the tallest bin in our localized histograms on each side of the mode since we cannot assume our posterior modes are symmetric, and then determine the range that spans five half-widths to half maximum on either side, which preliminary tests suggest is wide enough to encompass the vast majority of samples in each mode. Once this range  is determined, we create a final histogram for both the entire distribution as well as each mode (as seen in Figure \ref{fig:1388_posterior}) and select the center of the tallest bin as the maximum of that mode. The total probability of a mode is determined by integrating the marginalized posterior probability density within the five standard deviation range.


\subsection{3200 Phaethon}\label{sec:3200phaethon}

\begin{figure*}
    \centering
    \includegraphics[width=\textwidth]{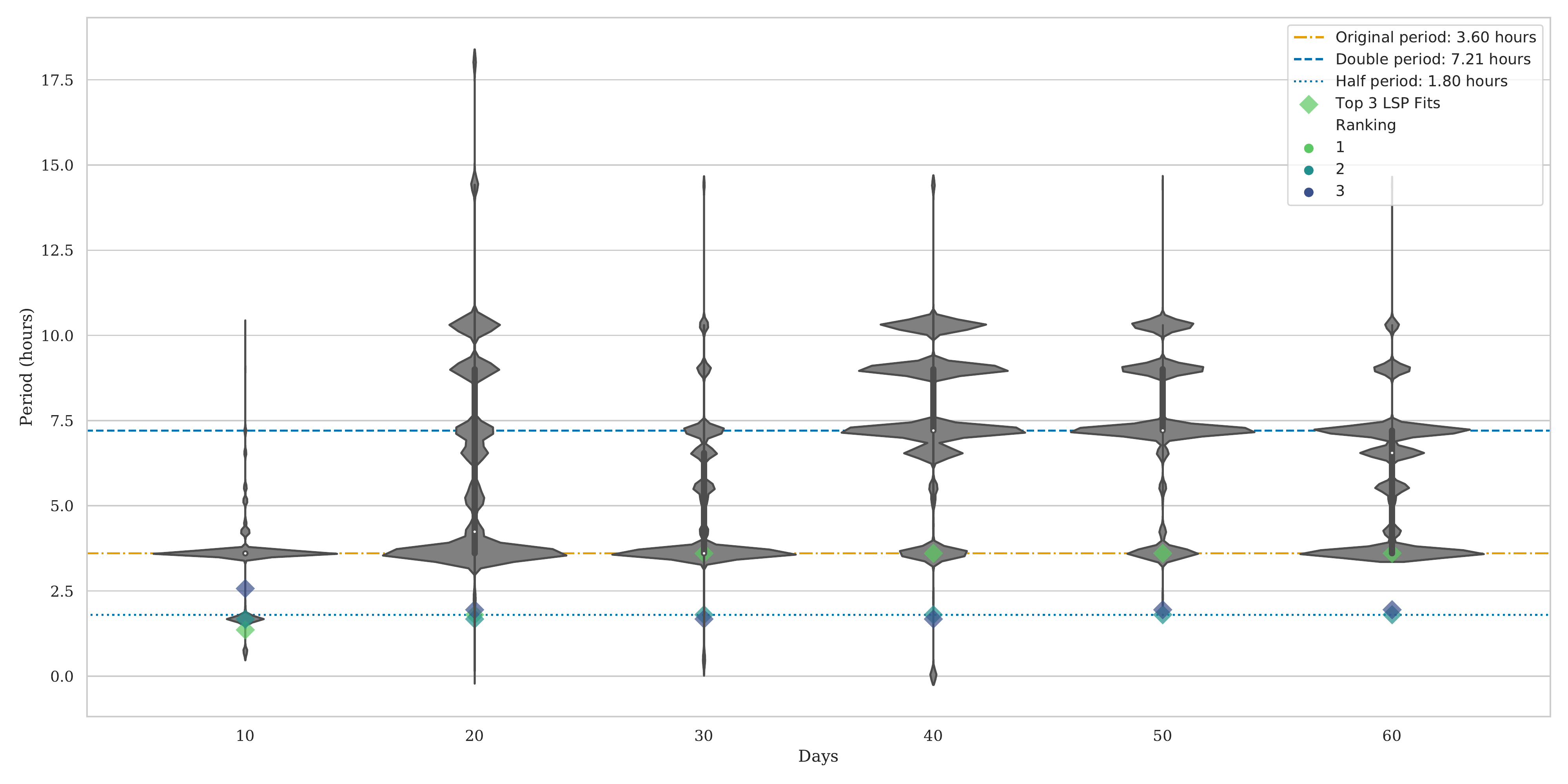}
    \caption{Violin plot (grey) showing posterior distributions of the different Gaussian Process results for Asteroid 3200 Phaethon, spanning 20, 60, 80, 100, and 120 data points (or 10, 30, 40, 50, and 60 days). Diamonds in green, teal and blue represent the top three best period estimates of Lomb-Scargle periodograms, respectively. Lines mark the period that was used to simulate the data (yellow, dot-dashed), double that period (blue, dashed) and half that period (blue, dotted).}
    \label{fig:3200_summary}
\end{figure*}

We simulate light curve data of Phaethon 3200, an Apollo-type near-Earth asteroid discovered in 1983 \citep{3200discovery}. Phaethon has an observed rotational period of 3.604 hours \citep{Phatheon} and a non-sinusoidal rotational profile (Figure \ref{fig:3200_unfolded}, right panel). We sub-sample the dense light curve generated from the 3200 Phaethon model in the DAMIT database with the ZTF-LSST cadence to generate individual realizations that vary in length from 10 to 60 days. With 2 observations per night, the generated light curves have between 20 to 120 data points.

We model each light curve with the Gaussian Process as described in Section \ref{sec:sim_setup} in order to infer the period and other parameters. The posterior probability distributions for a majority of the light curves are multimodal (Figure \ref{fig:3200_summary}), with modes at multiples of the true period, most notably twice the true period (7.208 hours) and 2.5 times the true period (9.010 hours). A significant fraction of the probability from the marginalized posterior distributions of the period for light curves with 10, 20, and 30 days of observations is located in a narrow mode around the \com{true} period (see Section \ref{sec:sim_setup}), 73.3\%, 46.2\%, and 57.8\% for 10, 20, and 30 days, respectively, while the light curves of 40 and 50 day durations have the majority of their probability mass located at double the \com{true} period, 32.5\% and  43.7\%, respectively. The light curve with 60 days of observations has its probability primarily split between the true period and double: 39.2\% and 30.4\%, respectively.
 
We generate LSPs for all simulated light curves and identify the three highest peaks in each (Figure \ref{fig:3200_summary}, diamonds). For light curves with 30 or more days of observations, the LSP estimates the true period and half the true period as the highest and second-highest peaks, respectively. However with 10 or 20 days of observations, the LSP fails to reliably identify the true period. The highest periodogram peaks for the 10-day light curve differ by more than 7 minutes from the true period. This is enough of a difference for a short-period asteroid like 3200 Phaeton with a period of a few hours to produce a folded light curve that appears incoherent. For the 20-day light curve, the highest peak suggests a period that is half of the true period.

\begin{figure*}
    \centering
    \includegraphics[width=\textwidth]{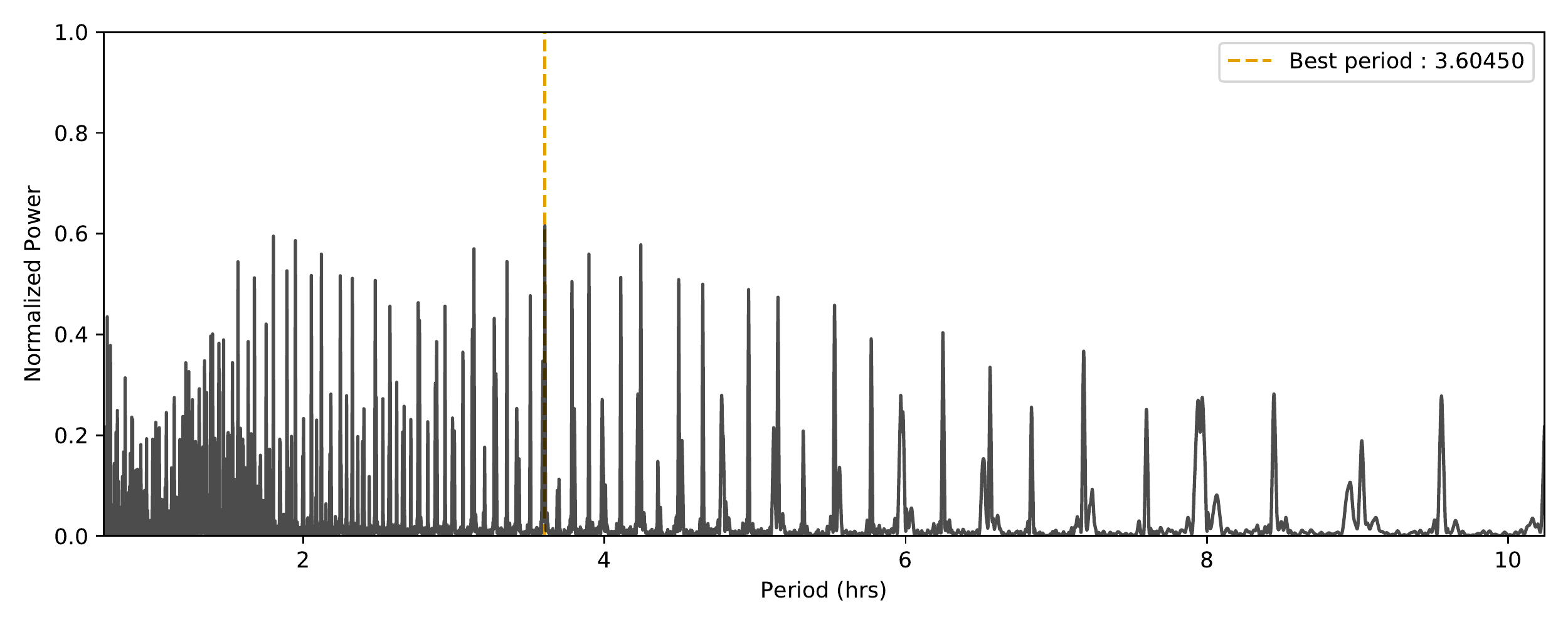}
    \caption{LSP for asteroid 3200 Phaethon for 60 days of observations ($N=120$ observations). The dashed yellow line indicates the highest peak in the periodogram, similar to the period found in \citet{Phatheon}.} 
    \label{fig:3200_60day_lsp}
\end{figure*}

For a better comparison between the LSP and the Gaussian Process posteriors, we also present the LSP for a simulated light curve over 60 days of observations in Figure \ref{fig:3200_60day_lsp}. The periodogram characterizes the best fit period within 2 seconds of the reported period \citep{Phatheon}, but also features several nearby peaks of similar heights.

\begin{figure*}
    \centering
    \includegraphics[width=\textwidth]{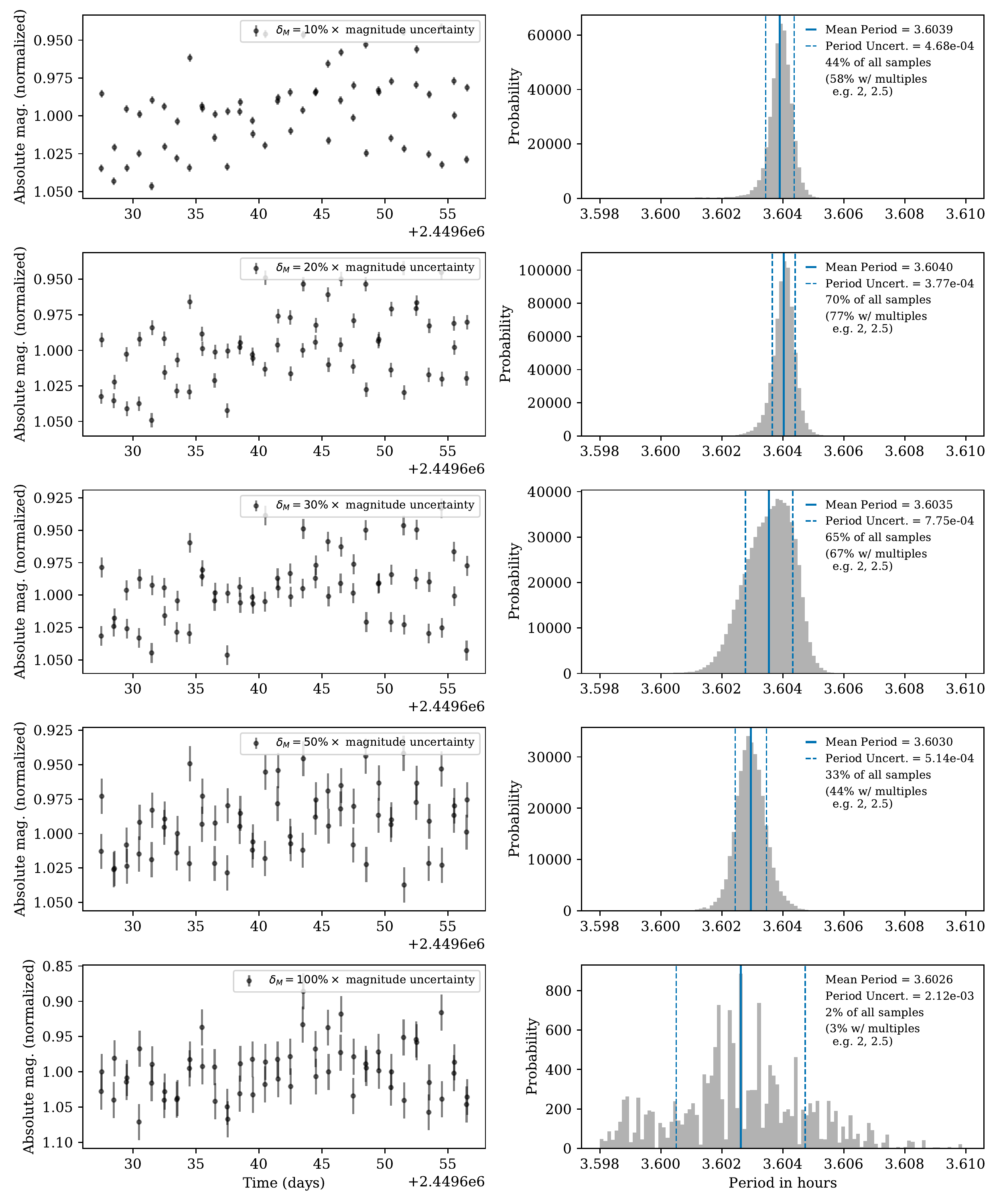}
    \caption{\com{Left: 30 days (60 data points) of simulated observations of asteroid 3200 Phaethon with varying magnitude measurements and uncertainties. Uncertainties are fractions of the standard deviation of the distribution of magnitudes, ranging from 10\% to 100\%. For each case, magnitude measurements are varied by sampling from normal distributions with $\mu$ equal to the original magnitudes and $\sigma$ equal to the magnitude error, respectively. Right: The marginalized posterior distribution of the period parameter near the true period, 3.604 hours. The solid blue line indicates the mean of the samples, and the dashed lines indicate $\pm 1\sigma$ of the distribution. The full marginalized posterior distributions (Figure \ref{fig:3200_variable_full}) can be found in the appendix in Section \ref{sec:appendix}.}
} 
    \label{fig:3200_variable}
\end{figure*}

\com{To test how the model performs in lower signal-to-noise (S/N) scenarios, we systematically increase the magnitude uncertainties and resample the magnitudes in the 30-day light curve of 3200 Phaethon (Figure \ref{fig:3200_variable}). This simulated light has 60 data points, and we increase our magnitude uncertainties to 20\%, 30\%, 50\%, and 100\% of the standard deviation of the light curve (for comparison, our other simulations use a magnitude uncertainty of 10\% of the standard deviation of the magnitudes). For each light curve, we vary magnitude measurements by sampling from normal distributions with $\mu$ equal to the original magnitudes and $\sigma$ equal to the magnitude error of each case, respectively. Parametrizing the magnitude error in terms of the standard variation allows us to test the effect of increasing the magnitude error compared to the intrinsic amplitude of the light curve on our statistical inference.}

\begin{figure*}[h]
    \centering
    \includegraphics[width=6.5in]{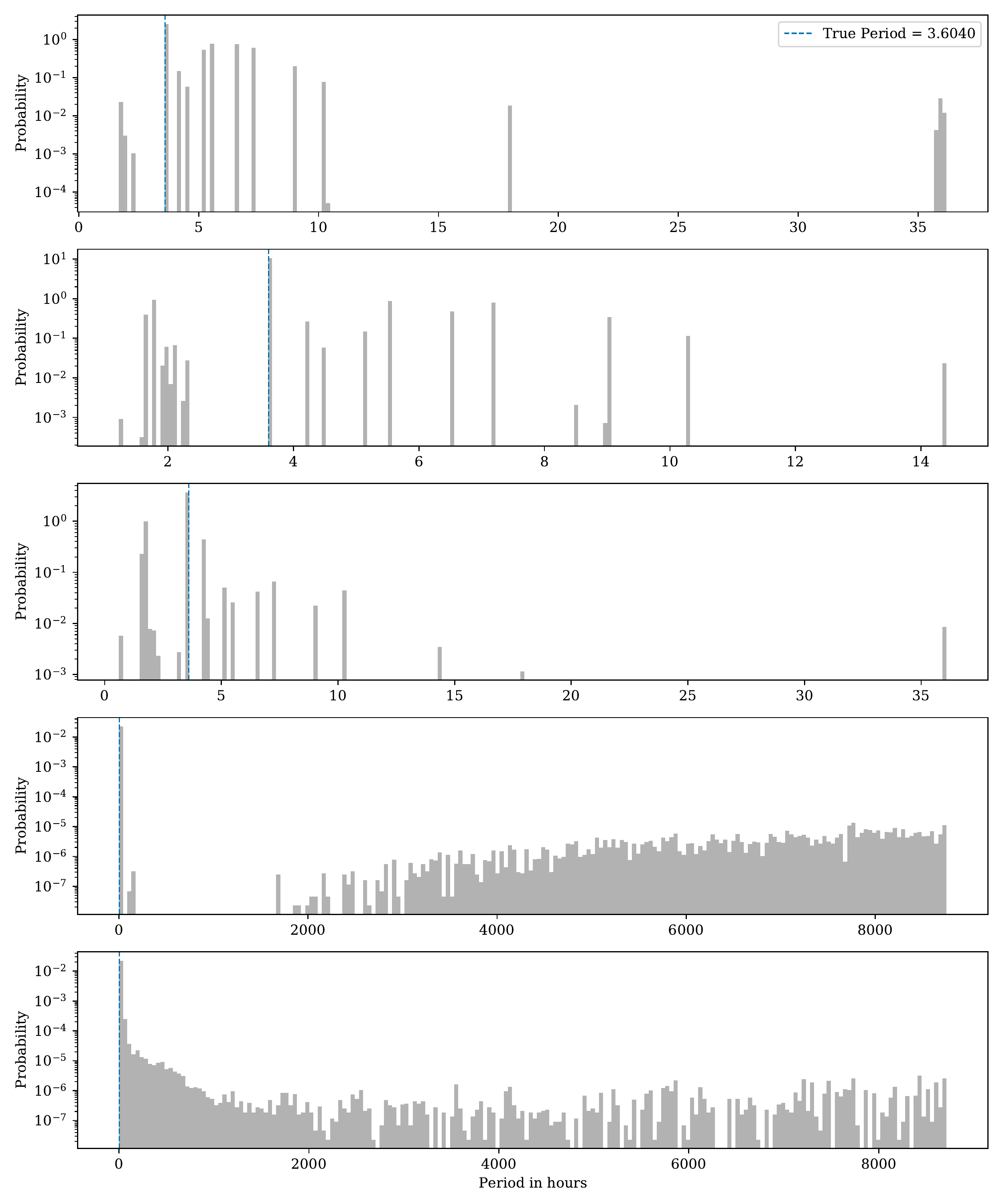}
    \caption{\textbf{The full marginalized posterior distribution of the period parameter with varying magnitude uncertainties, as opposed to Figure \ref{fig:3200_variable} which only shows the distribution near the true period. The dashed blue line indicates the true period from our simulated data.}
} 
    \label{fig:3200_variable_full}
\end{figure*}

\com{We find that in all test cases with less than 50\% magnitude uncertainty, the Gaussian Process successfully models the light curves. For these light curves, we find the largest mode of probability mass centered on the true period, constituting anywhere from 44\% to 70\% of all the probability. When multiples of the period (harmonics) are included, the summed probabilities range from 58\% to 77\%. For the case with 50\% magnitude uncertainty, we still find the largest mode of the probability mass centered on the true period (33\%, 44\% including harmonics), but the samples remain largely unconverged, as evidenced by the full marginalized posterior distribution (Figure \ref{fig:3200_variable_full}). Meanwhile, the Gaussian Process fails to model the case with 100\% magnitude uncertainty.}

\com{These results indicate that we are unable to model asteroids if their uncertainties are larger than $\sim$30\% of the standard deviation of their magnitude distribution. We report similar findings in Section \ref{sec:limitations} in regards to ZTF observations of 821 Fanny, which has an average magnitude uncertainty of $\sim$25\%. Given these results, we are optimistic that our model will be able to model objects with moderately low S/N.}


\subsubsection{Shape Determination via Amplitude Posteriors}

\com{We can use the posterior distributions to explore other relevant quantities to asteroid characterization. While the amplitude parameter is not directly related to the amplitude of the light curve itself, we can use posterior realizations to explore the amplitude variations. Assuming that 3200 Phaethon can be well-described as a tri-axial ellipsoid, we can use the amplitude variations to derive the axis ratio. To do so, we draw 1000 parameter sets from the posterior probability density. We use these parameters to calculate the Gaussian Process conditioned on the observations for a fine-grained grid of 1500 points over 1.5 days. We use these posterior draws to estimate the distance between maximum and minimum amplitude, taking into account both model and observational uncertainties. For this proof-of-concept, we use a simplified version of Equation (6) in \citet{lacerda2008photometric}, where we assume that the difference in amplitude $\Delta m$ due to the ratio in axes $a$ and $b$ for a tri-axial object rotating about axis $c$ can be described by}

\[
\Delta m = 2.5 \log(a/b) \; .
\]

\com{Here, we assume that $a=c$ and that attenuation of $\Delta m$ due to the solar phase angle and the aspect angle between the observer's line of sight and the rotational axis can be neglected.}

\begin{figure}
    \centering
    \includegraphics[width=0.5\textwidth]{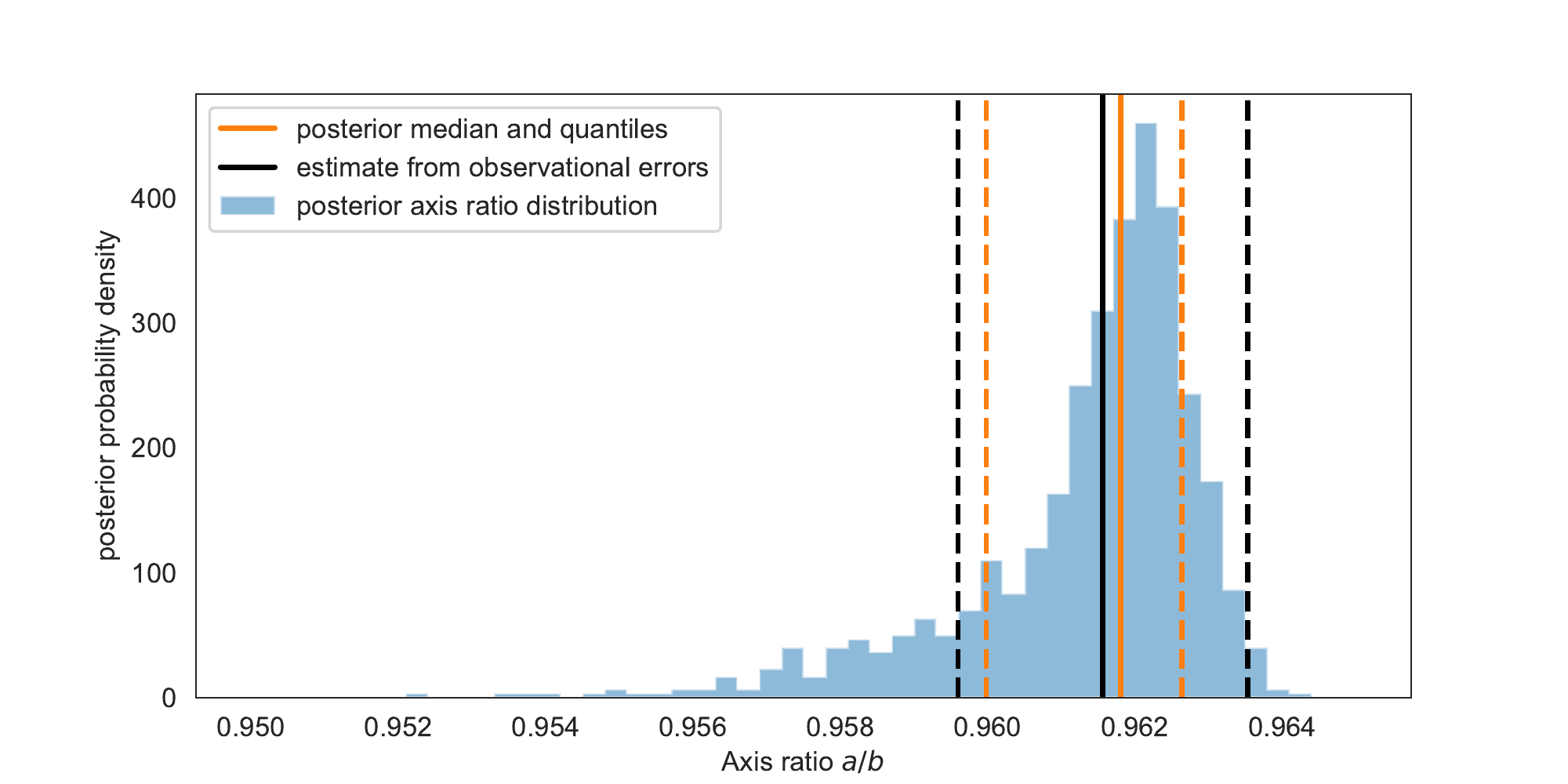}
    \caption{\com{Posterior probability density (blue) for the asteroid's axis ratio, based on the Gaussian Process model. For comparison, we show the standard estimate (solid black line) from the maximum/minimum data points and their observational uncertainties (dashed black lines).}}
    \label{fig:amplitude_shape}
\end{figure}

\com{Our results are reported in Figure \ref{fig:amplitude_shape}. The posterior probability density for the axis ratio is asymmetric towards smaller axis ratio values. It broadly agrees with a standard estimate from the maximum/minimum observed data points and their observational uncertainties. However, especially in sparse observations, it is not always entirely clear whether the observations sample the full extent in amplitude of the asteroid's light curve. Here, the Gaussian Process model can help improve our inference in two ways. Firstly, it can fold in prior knowledge about the asteroid itself where it exists, or about asteroid light curves more generally. Secondly, its estimate will take into account the properties of the asteroid based on the remaining data points and the small-scale structure inferred from them.}

\com{The axis ratio for 3200 Phaethon is not well-known. \citet{prsa2019} assume an oblateness of $a/b = 0.889$, similar to 
(101955) Bennu, broadly in agreement with radar measurements reporting a roughly spherical object \citep{2019P&SS..167....1T}. The latter conclusion is also in line with our own results of an approximately spherical object based on the posterior distributions and the simulated light curve derived from a shape model.}

\subsection{221 Eos}\label{sec:221eos}

\begin{figure*}
\includegraphics[width=\textwidth]{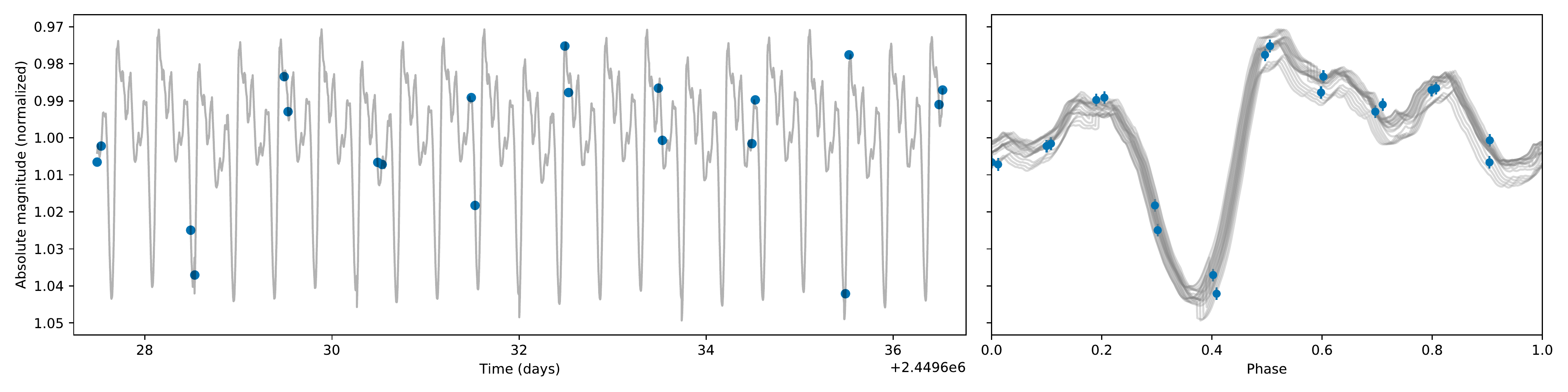}
\centering
\caption{Simulated light curve of asteroid Eos 221 over 10 days. Blue markers are 20 simulated observations and the grey line is the original simulated light curve obtained from the DAMIT database.}
\label{fig:221_lc}
\end{figure*}

We also test our model on 221 Eos, a well-documented main-belt asteroid with an average rotation period of 10.443 hours \citep{Eos}. Its light curve is highly structured, with a rotational profile departing significantly from a sinusoidal shape, as shown in Figure \ref{fig:221_lc}. Just as for 3200 Phaethon, we sub-sample the dense light curve generated from the 221 Eos model in the DAMIT database with the ZTF-LSST cadence for 10 to 60 days.

\begin{figure*}
    \centering
    \includegraphics[width=\textwidth]{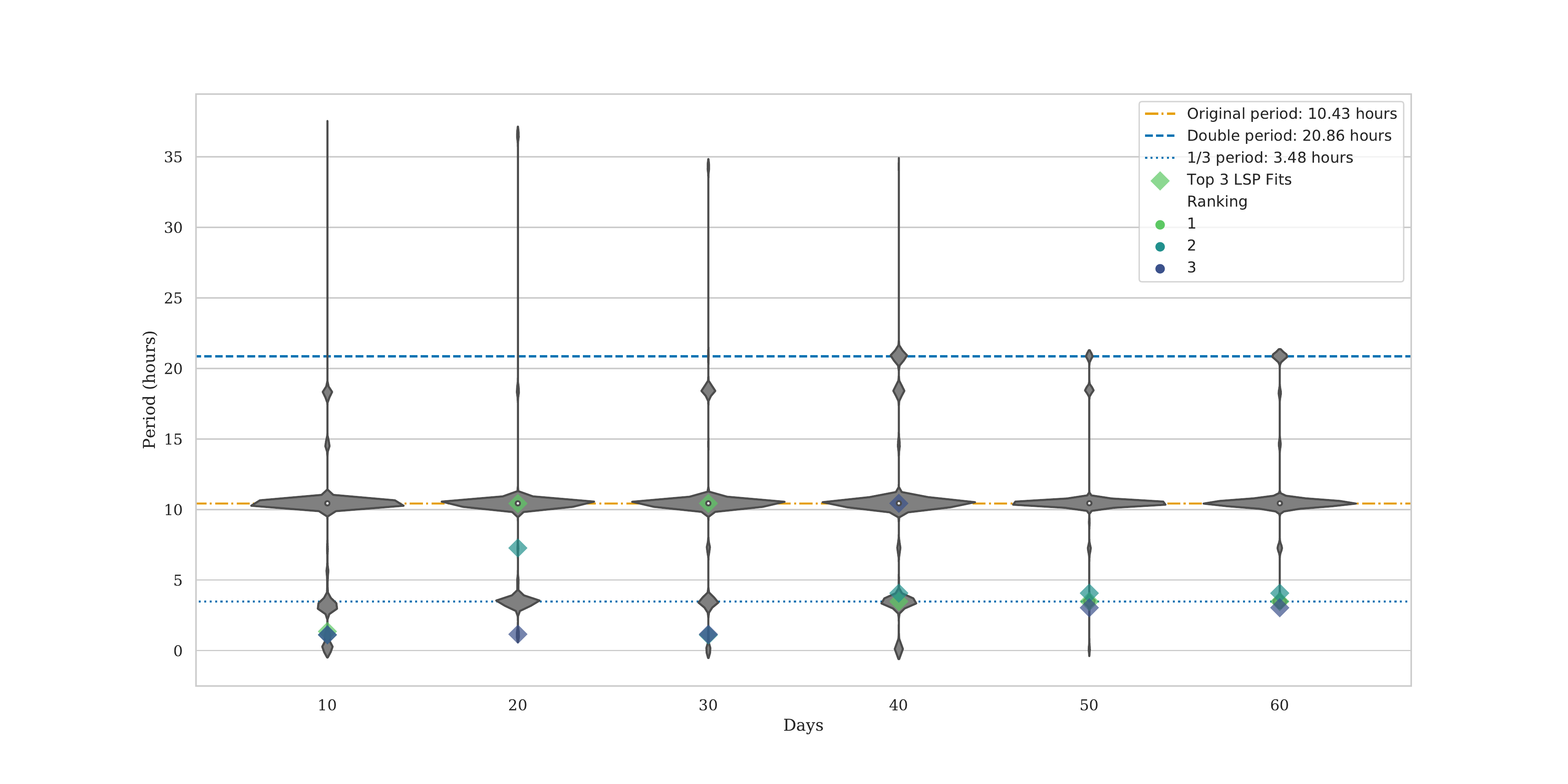}
    \caption{Violin plot (grey) showing posterior distributions of the different Gaussian Process results for Asteroid 221 Eos, spanning 10, 20, 30, 40, 50, and 60 days (20, 40, 60, 80, 100, and 120 data points). Colored diamonds in green, teal and blue represent the top three best period estimates in the Lomb-Scargle periodograms. The lines show the period that was used to simulate the data (yellow, dot-dashed), double (blue, dashed) that period, and 1/3 of the true period (blue, dotted). Our posterior probability distributions for the light curves with 30 and 40 days of observations contain a small number of samples more than five standard deviations (averaged across all samples) above the mean, which, for the sake of this visualization, have been masked.}
    \label{fig:221_summary}
\end{figure*}

We model each light curve with the Gaussian Process, as described in Section \ref{sec:sim_setup}. As for 3200 Phaeton, the resulting posterior probability distribution for each light curve is multimodal \textbf{(Figure \ref{fig:221_summary})}, with most modes at multiples of the true period. The Gaussian Process is highly effective in inferring the true period across all realizations and light curve durations. 
A large fraction of the probability from the marginalized posterior distributions of the period for each light curve is located in a narrow mode around the true period: 74.4\%, 74.8\%, 76.0\%, 64.4\%, 84.4\% and 80.5\% of samples for 10, 20, 30, 40, 50, and 60 days, respectively. There is also a smaller amount of probability located at a third of true period (\com{3.47} hours) for each of the light curves, possibly a reflection of the more complex substructure of this asteroid's rotational profile. 

Lomb-Scargle periodograms show a similar bias towards shorter periods in a more pronounced way: in many of the light curves, the highest peaks cluster around a third of the \com{true} period or less. For the light curve with only \com{10 days of observations}, none of the peaks were close to any obvious multiples of the true period while nearly 75\% of samples from the Gaussian Process model are clustered around the \com{true} period. Only for the light curves with 20 and 30 \com{days of observations} does the LSP show the \com{true} period as the highest peak.

\section{ZTF}\label{sec:ztf}

In order to understand the model's performance on real data, we test our model on observations from the Zwicky Transient Facility. We first look at a short-period main-belt asteroid to provide a direct comparison to our simulation results, then we focus on long-period asteroids for which period estimation is traditionally difficult.

\subsection{Data Reduction} \label{sec:ztf_data_reduc}

The ZTF alert stream provides aperture-corrected photometry for each object detected in difference images; detections near the predicted positions of known asteroids are tagged with this identification. We retrieved ZTF detections of a variety of asteroids from the public and private alert stream \citep{prsa2019}, removing any observations that fell more than 10 arcseconds away from the predicted positions of these (well-known) orbits. We also retrieved orbital information for each asteroid from the JPL Small Body Database (\url{https://ssd.jpl.nasa.gov/sbdb.cgi}) using Astroquery \citep{2019AJ....157...98G}. We used the orbital information to generated expected magnitudes at each epoch, using OpenOrb \citep{openorb}, thus providing a correction for distance and phase angle effects using a fairly simple $H-G$ phase curve (\cite{1989aste.conf..524B}) and assuming $G=0.15$. After subtracting these corrections from the observed magnitudes, we calculate the RMS of the resulting values to look for outliers more than $3\sigma$ away from the mean in each bandpass -- these outliers are typically the result of measurements compromised by a background star or edge effect and are rejected. We then calculate the mean corrected-photometric values per bandpass to get an initial estimate of the color of the asteroid; this color is also subtracted from the relevant photometric points. The resulting set of photometry for each asteroid is well corrected for distance effects, and approximately for phase curve effects and color. These `corrected' magnitudes center around zero in each and primarily reflect the effects of asteroid variability due to rotation.

Preliminary tests with the ZTF data show that the Gaussian Processes model breaks down in some cases, modeling photometric differences between bandpasses rather than finding the rotational period. This is likely driven by two problems which lead to an incorrect initial color estimate: inadequate modeling of the phase curve and uneven sampling of the rotational period in each bandpass. For some asteroids, using a single $G$ value with the $H-G$ phase curve correction is not sufficient, and offsets in the predicted magnitudes in different bandpasses at large phases lead to offsets in the estimated color. A more sophisticated phase curve model (and/or fitting $G$ values in each bandpass) could improve this. For other asteroids, the sparse sampling in each bandpass may fall more frequently into faint (or bright) points in the rotation period, again biasing the initial color estimate. 

In this initial study, where we primarily aim to show the use of this model on sparse observations, we proceed to model the data from $r$-band only. Future work will extend the model to allow observations from multiple bandpasses to be used simultaneously.
 
We initialize the MCMC samplers with the same setup as our simulated asteroids (see Section \ref{sec:sim_setup} for details) and determine the probabilities and peak periods for each mode in the posterior.

\subsection{Asteroid 1388 Aphrodite}

As a direct comparison to the asteroid light curves simulated in Section \ref{sec:sim}, we investigate a short-period main-belt asteroid from the Eos family. Asteroid 1388 was observed 121 times in the \textit{r}-band over the span of 139 days and has a reported period of 11.94389 hours \citep{1388aphrodite}.

\begin{figure*}
    \centering
    \includegraphics[width=13.5cm]{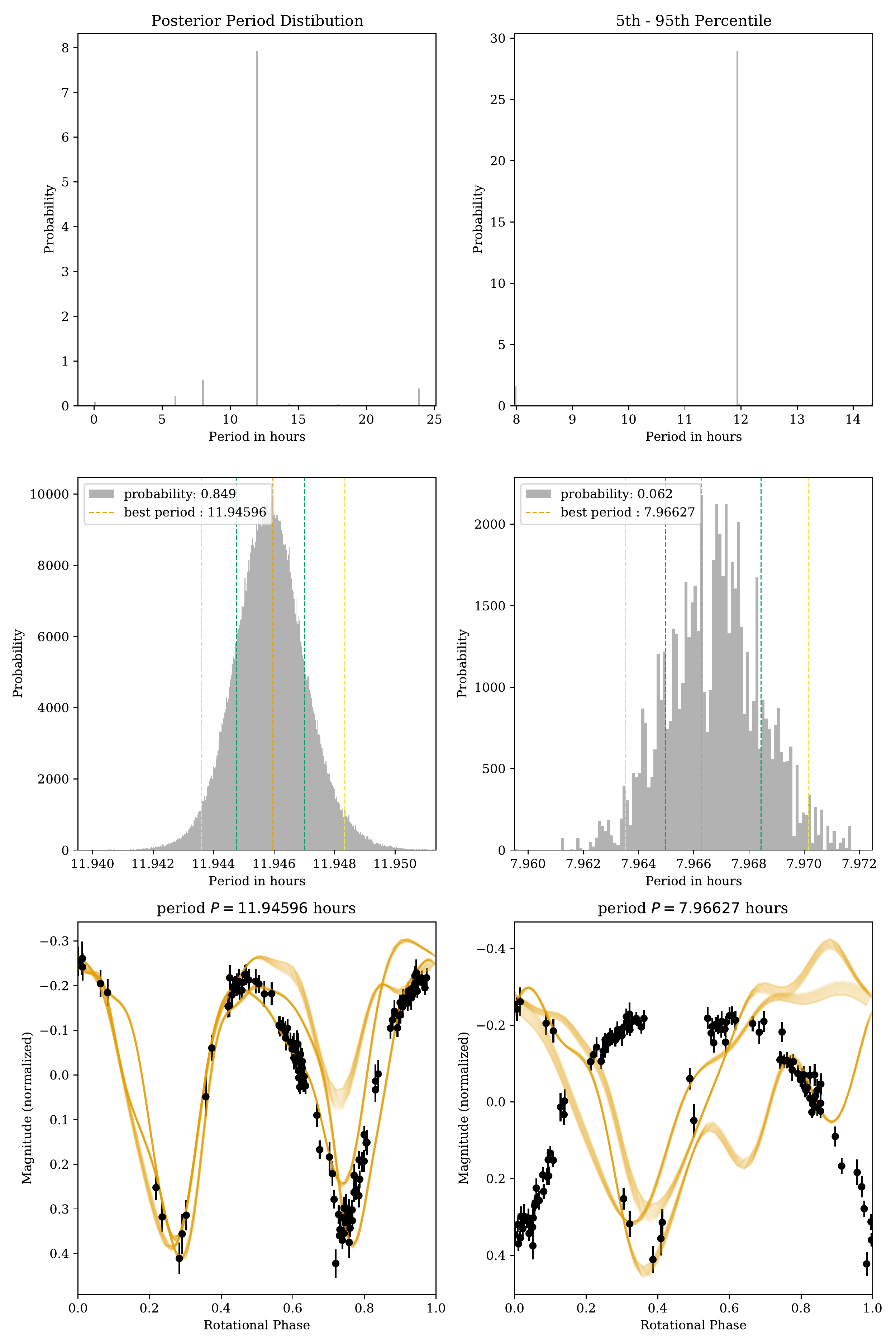}
    \caption{Top left: Posterior distribution of the period for Asteroid 1388 Aphrodite. Top right: 5th-95th percentile of the posterior distribution. Note that in the top two profiles, the posterior appears as thin, sparse peaks as a result of its multimodal topology. Middle: Modes with the highest and second highest probability mass (84.9\% and 6.2\% of samples, respectively), with the yellow line denoting the highest bin of the histogram, from which the period for the mode is calculated (11.94594 and 7.96627 hours). Dashed lines denote the 95th (light yellow) and 68th (green) percentile ranges. Bottom: ZTF observations (black) folded at the respective periods determined from the middle plots. Yellow lines are three realizations of Gaussian Processes with periods similar (within $\pm$ 0.5 hours) to the folding periods. These models span from the start of the observational window, to 20 times their respective folding periods, allowing us to see how the rotational profile varies over time.}
    \label{fig:1388_posterior}
\end{figure*}

\begin{figure}
    \centering
    \includegraphics[width=8cm]{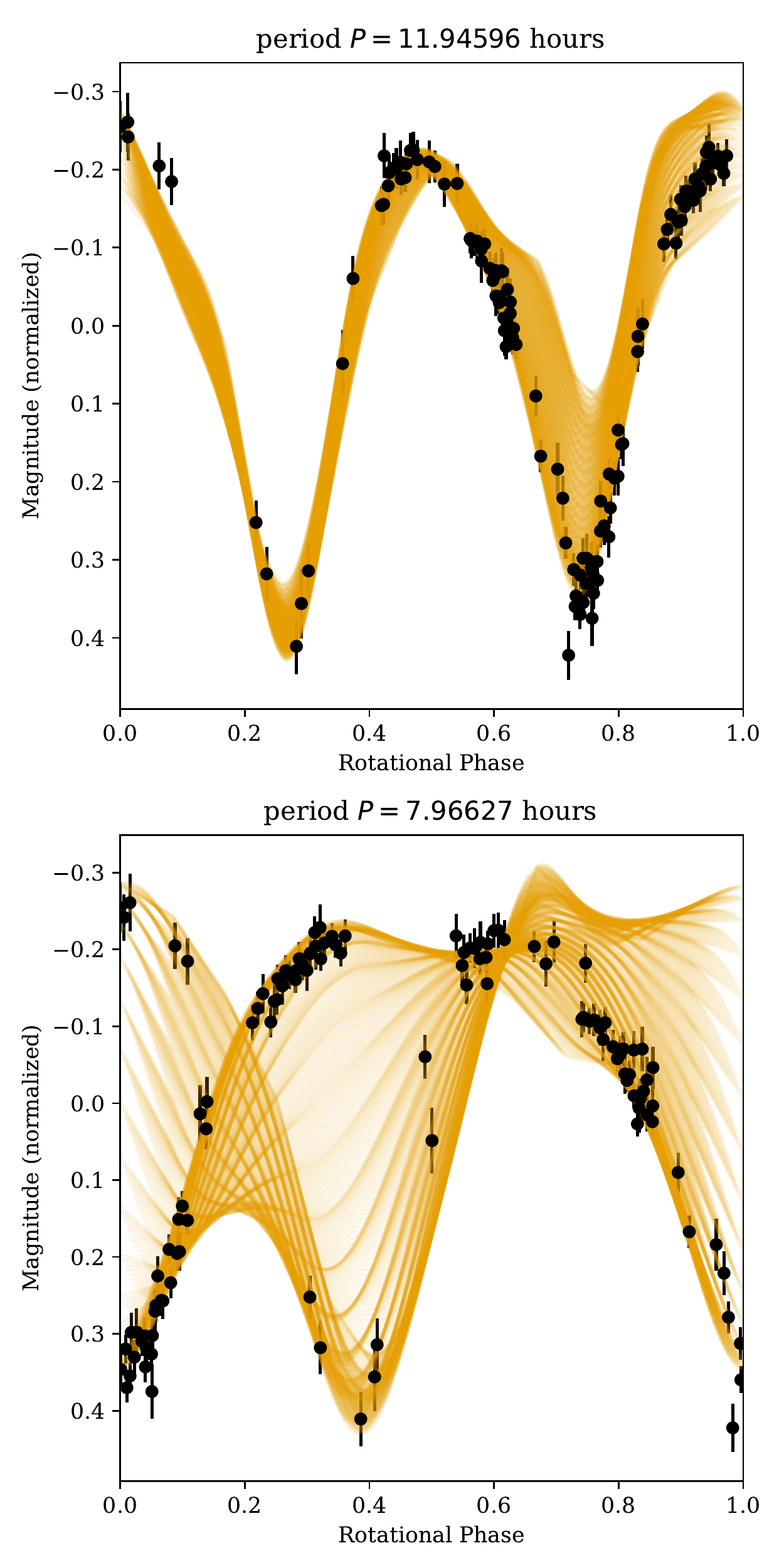}
    \caption{ZTF observations of asteroid 1388 Aphrodite folded at 11.946 hours (top) and 7.966 hours (bottom). The yellow lines are folded realizations of a Gaussian Process model conditioned on the data, and using parameters drawn from the mode at $11.946 \pm 0.5$ hours (top), and from the mode at $7.966 \pm 0.5$ hours. As opposed to the bottom panels in Figure \ref{fig:1388_posterior}, these models span the entire observational window of the asteroid. Covering the full observational range reveals how the rotational profile varies over time. This variation is particularly apparent in the bottom panel, where the model exploits a degeneracy between the rotational period and the metric $M$ to generate a light curve that matches the observations despite a shorter period $P$ by enforcing large changes in the rotational profile over the 139 days of observations.}
    \label{fig:1388_folded}
\end{figure}

\begin{figure*}
    \centering
    \includegraphics[width=\textwidth]{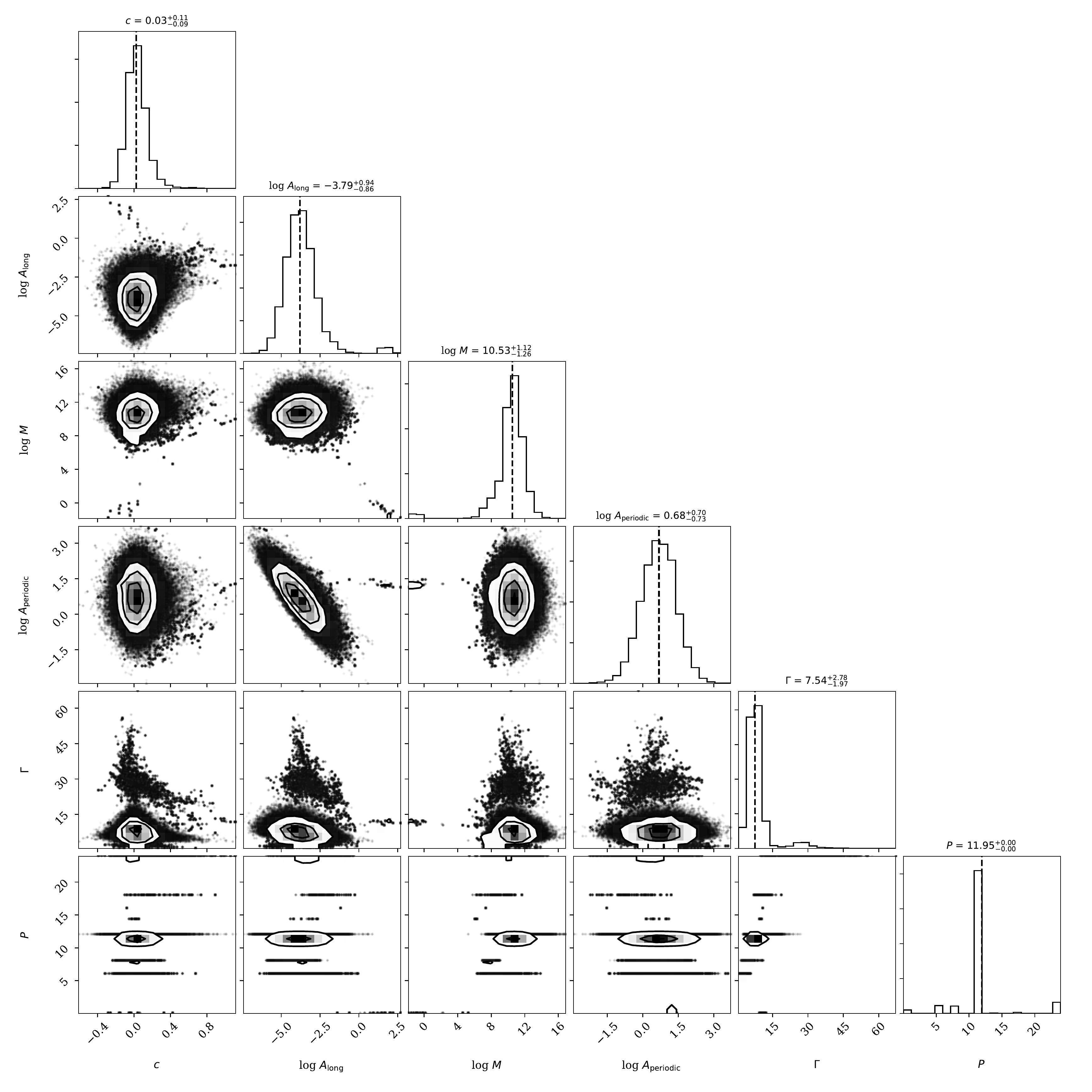}
    \caption{This figure, also called a \textit{corner plot}, presents a visualization of the posterior probability distribution of model parameters for Asteroid 1388 Aphrodite. Because the posterior pdf is a six-dimensional distribution, this visualization aims to make the information contained in the posterior accessible by unfolding it into a series of 1D and 2D representations. On the diagonal, we present histograms of the marginalized posterior probabilities for each parameter. On the off-diagonal, we show scatterplots for all pairwise combinations of parameters. The median of each distribution is marked on the histograms and included in the title of each column, along with the 16th and 84th percentile differences. $A_\mathrm{long}$, $M$, and  $A_\mathrm{periodic}$ are plotted on a logarithmic scale.}
    \label{fig:1388_corner}
\end{figure*}

We present our posterior inferences in Figure \ref{fig:1388_posterior} (summaries of the marginalized posterior of the period parameter, and folded light curves), Figure \ref{fig:1388_folded} (a more complete representation of the folded light curve with models), and Figure \ref{fig:1388_corner} (a corner plot summarizing posterior inferences for all parameters). 
84.9\% of the probability in the posterior distribution of the period are concentrated in a narrow mode centred on a period of 11.94596 hours (Figure \ref{fig:1388_posterior}). There is a smaller mode centred on the double of the period (23.89141 hours), although it only accounts for 4.1\% of the total posterior probability. Around 6.2\% of the probability mass is at 7.96627 hours, i.e. at 2/3 of the narrow posterior peak with the highest overall probability. Overall, over 95\% of the probability reflects an estimate close to or at a multiple of the previous reported period.

We plot our original observations folded at the periods reflecting the peak of each of the two most probable modes (11.95 and 7.97 hours) in the lower row of Figure \ref{fig:1388_posterior}. To visualize the Gaussian Process, we use parameter values from samples from the same mode in the marginalized period posterior distribution to construct realizations of the Gaussian Process for an interval of about twenty periods (240 and 160 hours, respectively) from the start of the observing window. We fold these model lightcurves at the same respective periods as mentioned above to illustrate the behaviour of the model. For the data folded at 11.95 hours, the light curve resembles a double-peaked sinusoid, which is common for asteroids shaped like tri-axial ellipsoids \citep{Harris2014}. 

The light curve folded at 7.97 hours does appear to have a visible pattern (Figure \ref{fig:1388_posterior}, bottom right panel), but the models for the first 160 hours reveal that the Gaussian Process appears to be mainly modelling a group of dimmer observations from the first dip of the double-peaked sinusoid while simultaneously modelling some of the brighter observations by varying the light curve slowly over time. Because we have opted to only plot the model light curves for the first 160 hours for clarity, we present in Figure \ref{fig:1388_folded} the same data and model, but spanning the full observational window rather than only the first 160 hours. The lower panel of this plot reveals a significant evolution in the intra-period variability over the 139 days of observations, where the peak shifts by nearly one half of a rotational phase in order to fit all data points. We discuss this behaviour further in Section \ref{sec:discus}.

We present a corner plot of the posterior distribution in Figure \ref{fig:1388_corner}.
As expected, the posterior probability of the constant mean flux $c$ is consistent with zero, $0.02 \pm 0.1$, indicating that the subtraction of the phase angle during pre-processing largely succeeded. The two-dimensional scatter plots reveal some scatter in the distribution, and a slight bimodality for the amplitude $A_\mathrm{long}$ and the metric parameter $M$ of the squared exponential covariance function, as well as the inverse length scale of the periodic kernel $\Gamma$. 

We observe a negative correlation in parameter space between the log-amplitude $A_\mathrm{long}$ for the long-term squared exponential covariance function and the log-amplitude for our periodic sine squared exponential covariance function, $A_\mathrm{periodic}$. This trend is likely a result of both parameters contributing to the overall amplitude of the combined covariance functions: if one amplitude has a high value, the other amplitude will need to be at a lower value to produce the overall observed variability.

There is a slight positive correlation between the inverse length scale $\Gamma$ of the periodic covariance function, which models the intra-period variability, and the period, $P$. This is expected: the longer the proposed period, the more complex the variations that the Gaussian Process has to model within each period. For example, a double sinusoid exhibits more complex intra-period variability when compared to a single sinusoid, so a longer period would correlate with a greater inverse length scale $\Gamma$ value.

\subsection{Asteroid 299 Thora}

Because our prior favours short periods, we want to explore the model’s performance on a long-period asteroid. Asteroid 299 Thora is a long-period main-belt asteroid with a reported period of 273 hours \citep{299thora}. It was observed 49 times by ZTF in the \textit{r}-band over 153 days.

\begin{figure*}
    \centering
    \includegraphics[width=13.5cm]{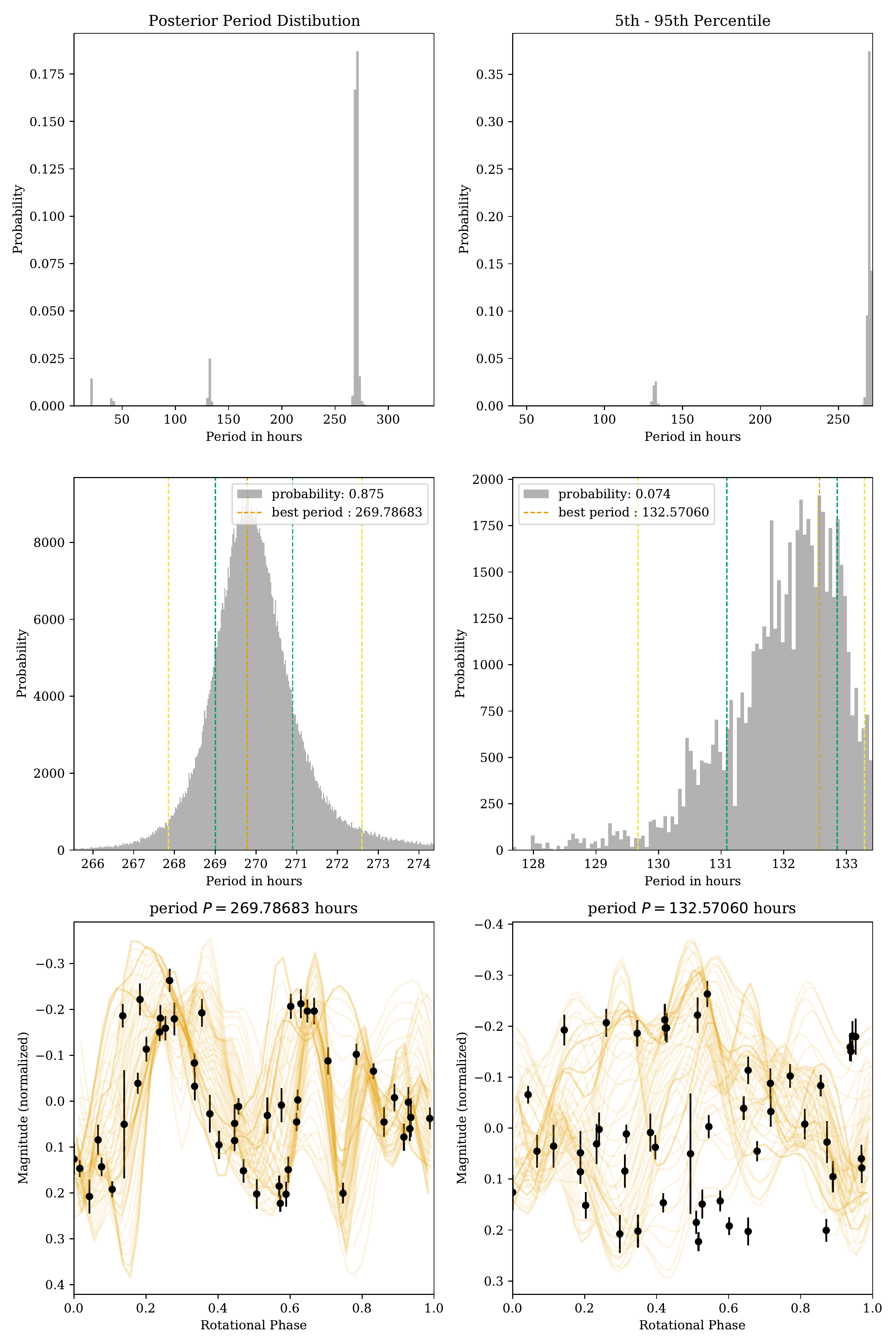}
    \caption{Top left: Posterior distribution of the period for Asteroid 299 Thora. Top right: 5th-95th percentile of the posterior distribution. Middle: Modes with the highest and second highest probability mass (87.5\% and 7.4\% of samples, respectively), with the yellow line denoting the highest bin of the histogram, from which the period for the mode is calculated (269.78683 and 132.57060 hours). Dashed lines denote the 95th (light yellow) and 68th (green) percentile ranges. Bottom: ZTF observations (black) folded at the respective periods determined from the middle plots. Yellow lines are three realizations of Gaussian Processes with periods similar (within $\pm$ 0.5 hours) to the folding periods. These models span from the start of the observational window, to 20 times their respective folding periods, allowing us to see how the rotational profile varies over time.}
    \label{fig:299_posterior}
\end{figure*}

The posterior probability distribution (Figure \ref{fig:299_posterior}) of the period indicates that over 87\% of the probability is located in a mode around a peak at 269.79 hours, though the posterior is asymmetric. The posterior has a second mode at 132.57 hours (approximately half of our first period), containing 7.4\% of the probability. 

\begin{figure*}
    \centering
    \includegraphics[width=\textwidth]{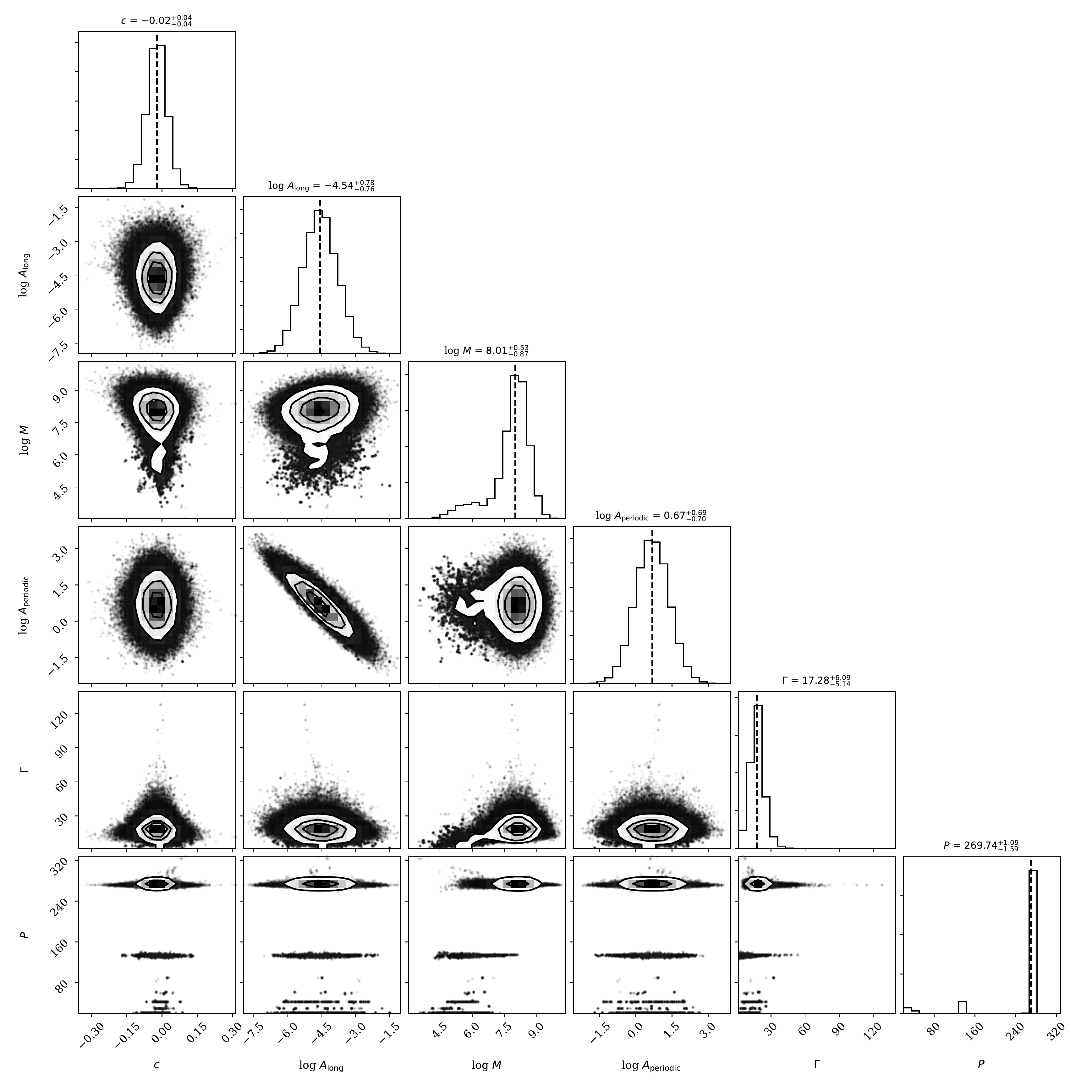}
    \caption{Corner plot of model parameters for Asteroid 299 Thora. The 50th percentile of each distribution is marked on the histograms and written in the title above, along with the 16th and 84th percentile differences. $A_\mathrm{long}$, $M$, and  $A_\mathrm{periodic}$ are plotted on a logarithmic scale.}
    \label{fig:299_corner}
\end{figure*}

The corner plot (Figure \ref{fig:299_corner}) once again shows a large negative correlation between the two amplitudes, $A_\mathrm{long}$ and $A_\mathrm{periodic}$, and a slight positive correlation between the log metric $M$ and inverse length scale $\Gamma$ parameters. All other parameters are well-constrained in a single mode.

\subsection{Asteroid 821 Fanny} \label{sec:821fanny}

We include the asteroid 821 Fanny as an illustration of the limits of the model. Asteroid 821 is a slowly-rotating main-belt asteroid with a reported period varying from $230.6 \pm 0.3$ hours to $236.6 \pm 0.3$ hours in one recent analysis \citep{821fanny230-236h} and $238.9 \pm 0.8$ hours in another \citep{821fanny239h}. Asteroid 821 was observed by ZTF 46 times in the \textit{r}-band over 153 days. We initially sampled the model for 11,000 iterations, but convergence criteria indicated that the chains had not converged, and thus continued to sample for another 10,000 iterations.

\begin{figure*}
    \centering
    \includegraphics[width=\textwidth]{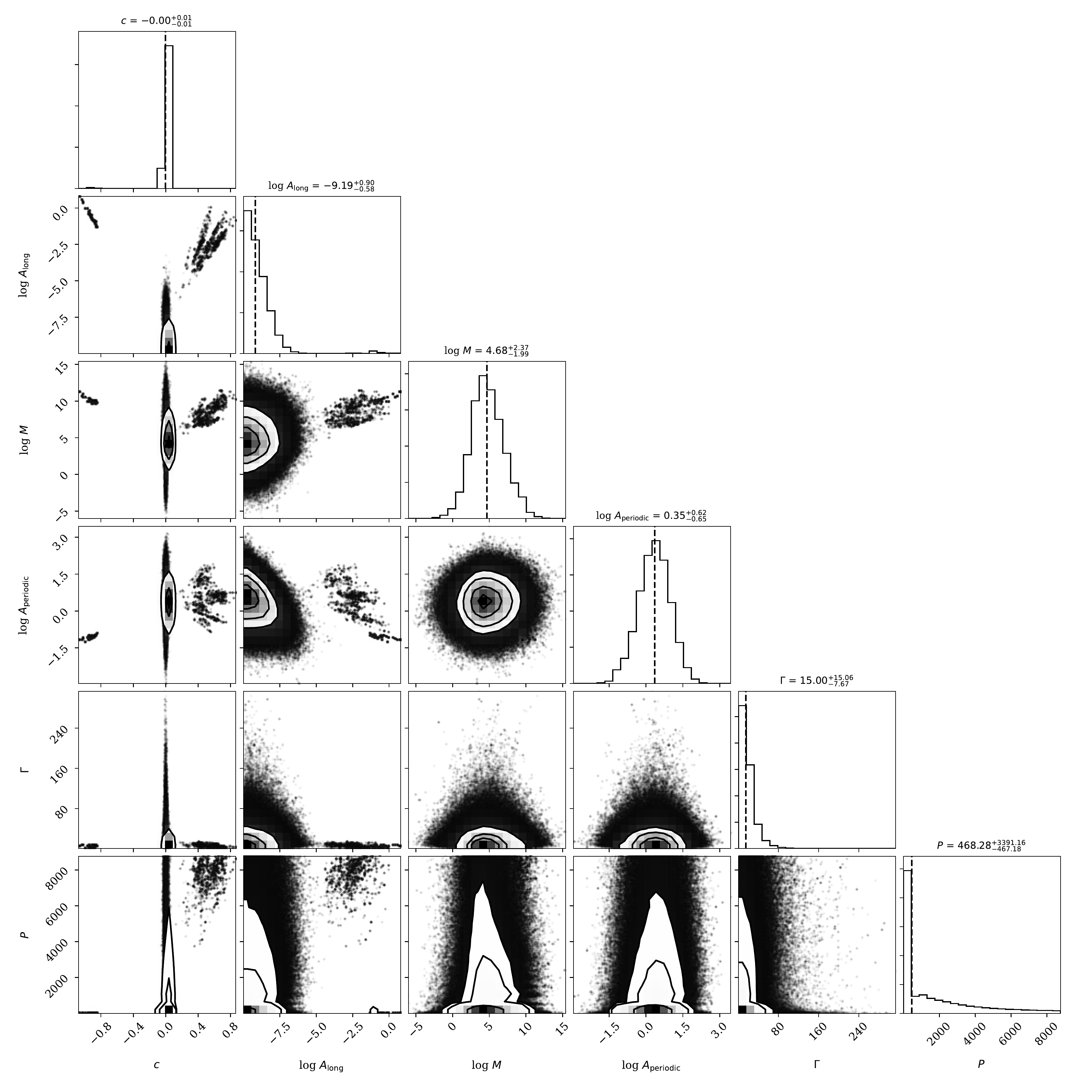}
    \caption{Corner plot of model parameters for Asteroid 821 Fanny. The 50th percentile of each distribution is marked on the histograms and written in the title above, along with the 16th and 84th percentile differences. $A_\mathrm{long}$, $M$, and  $A_\mathrm{periodic}$ are plotted on a logarithmic scale.}
    \label{fig:821_corner}
\end{figure*}

The period parameter spans the entire allowed prior range (Figure \ref{fig:821_corner}) as described in Section \ref{sec:priors}, suggesting that either the sampler has still not converged, or that the period is entirely unconstrained given the data. The inverse length scale $\Gamma$ also spans a very large range (values $\sim100-1000$) when compared to better-constrained asteroids discussed earlier. This is expected, given that, if the model suggested a period of several months (much longer than our total observing time for this asteroid), then the inverse length scale $\Gamma$ has to encapsulate all the variations in the flux, or the  intra-period variability, over the span of that period.

\begin{figure}
    \centering
    \includegraphics[width=8cm]{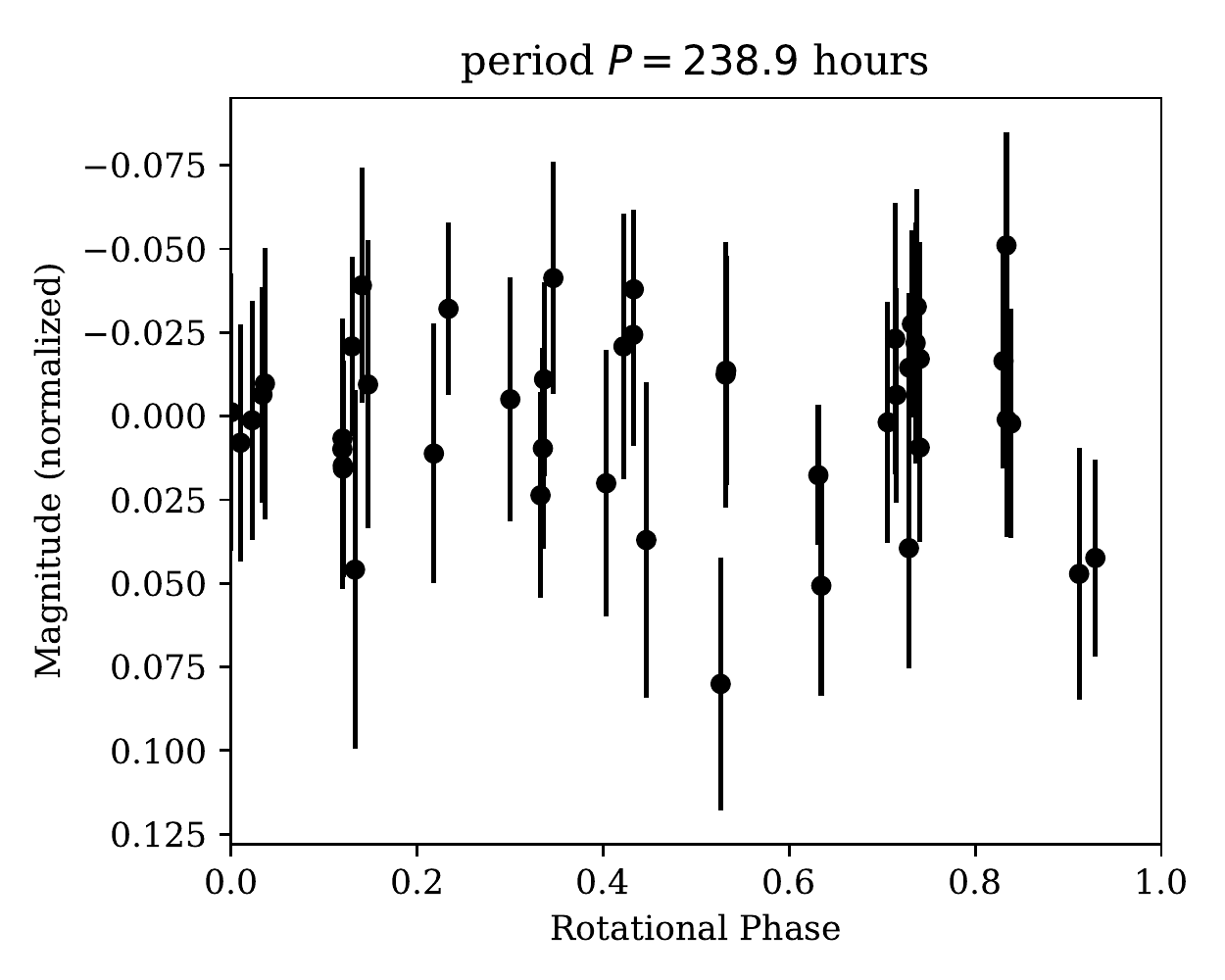}
    \caption{ZTF observations of Asteroid 821 Fanny folded at one of the reported periods, 238.9 hours.}
    \label{fig:821_reported}
\end{figure}

Folding the data to 236.6 hours, one of the periods reported by \citet{821fanny230-236h}, yields a noisy light curve (Figure \ref{fig:821_reported}) without much visible structure, but highlights the high magnitude uncertainty. In cases like this, where the amplitude due to rotation is smaller than the uncertainties on the data points and the data is sparse, any period yields an acceptable solution.

\begin{figure}
    \centering
    \includegraphics[width=8cm]{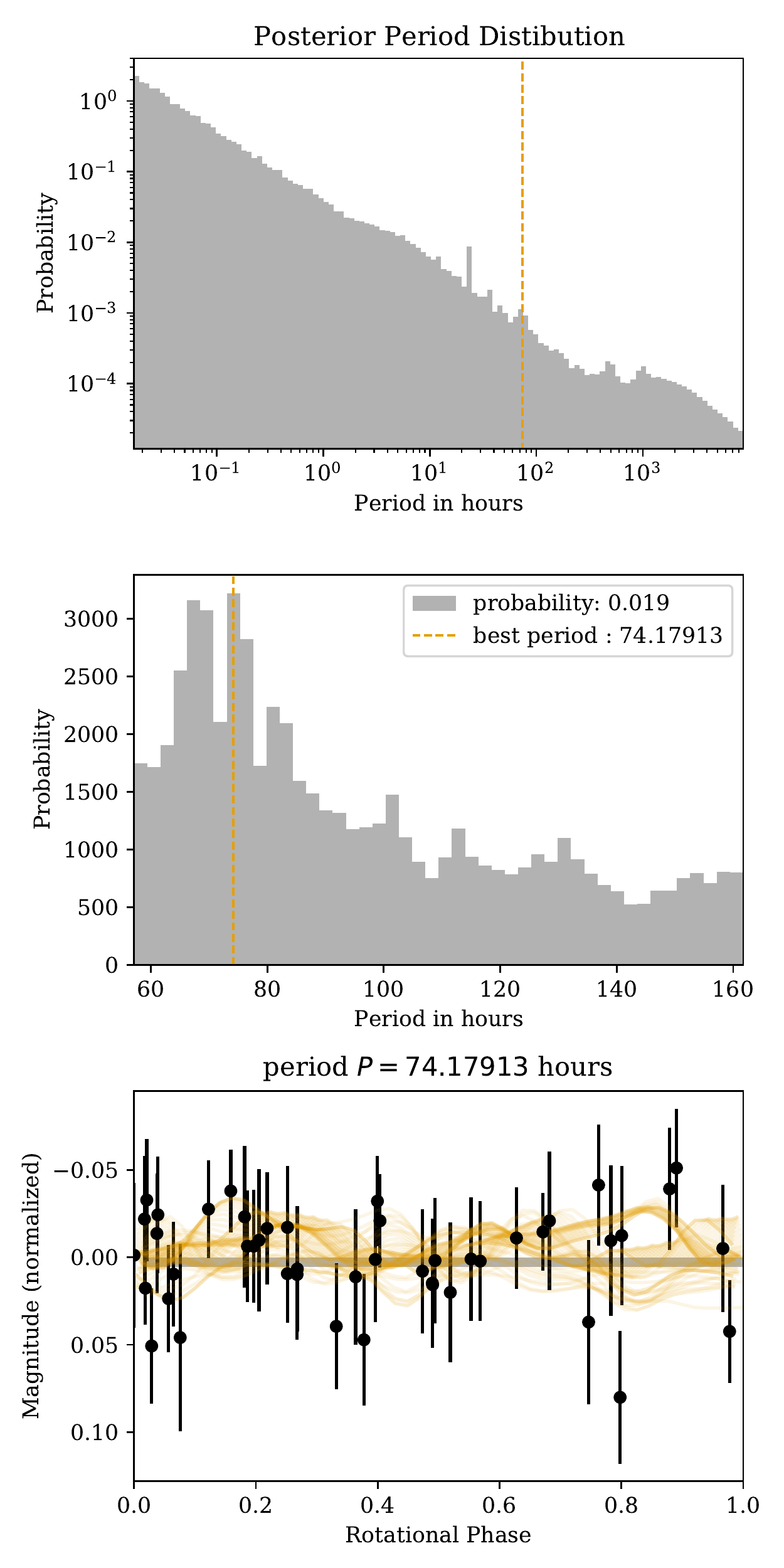}
    \caption{Top: Posterior distribution of the period for Asteroid 821 Fanny, with logarithmic scales on both axes with a yellow line denoting 74.18 hours. Middle: Mode with the highest probability mass (1.9\% of samples) with a yellow line denoting the highest bin of the histogram, from which the period for the mode is calculated (74.18 hours). Bottom: ZTF observations (black) folded at the period determined from the middle plot. Yellow lines are three realizations of the Gaussian Process with periods similar (within $\pm$ 0.5 hours) to the folding period. These models span from the start of the observational window, to 20 times the folding period (\~1500 hours).}
    \label{fig:821_posterior}
\end{figure}

The posterior distribution (Figure \ref{fig:821_posterior}) for the period parameter has some small peaks, but looks extremely broad overall, as expected from the corner plot (Figure \ref{fig:821_corner}). That the posterior probability peaks at the lower edge of the prior could in principle indicate that the solution that best represents the data sits beyond the 1-minute cutoff specified in the prior. However, smaller periods are believed to be unphysical, since they would yield an asteroid that spins itself apart \citep{Jewitt2017, LCDB, Chang2015, Moreno2017}. The minor peak that does exist in the probability distribution is located around 74.18 hours. We fold the observations and generated models for 74.18 hours and find model light curves with very small amplitudes, broadly consistent with a constant value. It is also possible that our model did not converge on a clear period if this asteroid is in a non-principal axis rotation state.

\section{Discussion} \label{sec:discus}


The standard model for an asteroid--a tri-axial ellipsoid--produces a double-peaked light curve where each peak is approximately sinusoidal. In practice, however, asteroids display a wide range of rotational profiles that deviate from that simple model. Irregular rotational profiles are observed when asteroids have either patches of surface materials with different albedos, abnormal shapes, or are relatively spherical but have surface variations that affect reflectance. 
The resulting light curves can be complex and generally exhibit multiple peaks as well as other substructure. In these cases, the LSP requires a large ensemble of sinusoidal functions to accurately represent the light curve, leading to power spread over many frequencies. This in turn complicates the inference of the underlying period. The behaviour is evident from the two simulated asteroids, particularly 221 Eos (Figure \ref{fig:221_summary}), where much of the power in the LSP is concentrated near 1/3 of the true rotational period. This can be explained by 221 Eos' triple-peaked structure (Figure \ref{fig:221_lc}).

In contrast, the flexibility of the Gaussian Process model implemented admits a wide range of realistic light curve shapes observed in nature. The sine squared exponential covariance function only assumes an exponential decay of the covariance of data points within a period $P$, meaning the Gaussian Process has the flexibility to model a wide range of realistic rotational profiles as long as they remain periodic (and do not drastically change over a short time; Figure \ref{fig:1388_folded} shows how rotational profiles could potentially morph over time). In addition, we include another covariance term to account for long-term changes in the rotational profile as a function of time. The squared exponential covariance function enables the model to account for the natural evolution of a rotational profile using the log-metric parameter $M$, as the phase-angle gradually changes throughout an asteroid's orbit. We expect asteroid lightcurves to evolve over longer periods of time as the phase-angle between us, the asteroid, and the sun changes. This is especially the case for sparse asteroid light curves, where data collection routinely spans months to years, and occasionally decades.

These additional degrees of freedom allow the model to better characterize sparse data obtained over a large range of time, along with differing rates of phase-angle change.  In the future, we will implement other physical processes--the phase angle via a mean function, tumbling asteroids via additional terms in the covariance function--to allow for realistic modeling of a wider range of asteroids.

The model implemented is a Bayesian model: this allows straightforward inference of relevant model parameters through the posterior distribution, as well as incorporation of known asteroid properties through the prior probability distributions.  The model allows for principled \textit{population inference} through Bayesian hierarchical modeling. While in this paper, we focused on the inference of properties of individual asteroids, hierarchical models allow for the joint inference of both individual properties as well as population parameters (e.g. the distribution of periods, amplitudes--as a proxy for shape--and intra-period variability) for physically related groups of asteroids. For example, the empirical prior for the period introduced in Section \ref{sec:priors} could then be replaced by a physically motivated distribution, and the population-level parameters of this distribution be inferred.


The model performs well at characterizing simulated light curves of asteroids, especially for sparse observations with a small number of data points, as long as the intrinsic variance in the magnitude due to rotation exceeds the measurement uncertainties. The posterior probability for both simulated and observed light curves is multi-modal, but concentrated in a small number of narrow peaks, compared to the wider spread in power across frequencies in the Lomb-Scargle periodogram (Figure \ref{fig:3200_60day_lsp}). The majority of the probability density is concentrated for each simulated light curve at either the true period or at twice that value. 

The posterior probability for the 3200 Phaethon simulation with the fewest observations (10 nights and 20 data points, see Figure \ref{fig:3200_summary}) has its highest mode at the true period (73\%). This simulation is also the only 3200 Phaethon light curve for which the LSP fails to find either the true period or any of its multiples within its three highest peaks.


Overall, the model performs well when data sets span multiple rotational periods, and the magnitude uncertainty is relatively small. Failure to meet these criteria results in incorrect parameter inference (see also Section \ref{sec:limitations}). We report to a high degree of confidence that the rotational period for Asteroid 1388 Aphrodite is 11.94597 hours with a 68\% credible interval ranging from 11.94475 to 11.94700 hours. The previously observed rotational period of 11.94389 \citep{1388aphrodite} falls within a 95\% credible interval. \citet{1388aphrodite} uses 16 dense light curves along with an additional 220 sparse individual measurements to calculate their period. We use 121 observations spaced over 139 days to arrive at a compatible result. 

Asteroid 1388 Aphrodite has a secondary mode in its posterior probability distribution of the period with a peak $P$ of 7.97 hours and an $M$ value of 1754.6 days based on Figure \ref{fig:1388_corner} (as opposed to an $M$ value of 45251.9 days for the highest probability mode with $P$ 11.95 hours). This means the Gaussian Process models variations in the light curve and phase-angle on a much shorter timescale at 7.95 hours than at 11.95 hours. In practice, while both periods presented are possible in theory, the posterior distribution for the period $P$ favors the previously established period of 11.94389 hours. The evolution of the rotational profile can be seen directly in Figure \ref{fig:1388_folded}, where a realization of the Gaussian Process model with a period near 7.97 hours (yellow) has been plotted and folded for the entire observational window. 

While this secondary mode is less probable than the primary mode at $P$ 11.97 hours (and therefore likely is not the correct intrinsic period, especially when compared to previously-reported periods), it illustrates how in the limit of sparse observations, there is an intrinsic degeneracy between the period $P$ and the metric $M$ of the squared exponential covariance function: in order to accommodate a period that is shorter than the likely underlying period, the model adjusts to incorporate rapid changes in the rotational profile in order to adequately represent all data points. Whether this is realistic or not depends on the actual location and movement of the asteroid, something we currently do not directly take into consideration. For distinct populations of asteroids with some known physical properties (e.i. near-Earth asteroids versus Trojan asteroids), it may be advisable to modify the prior so that the metric parameter $M$ better reflects the timescales over which we might expect the shape of the rotational profile to change. Here, we use a relatively broad prior for $M$ to model asteroids since we are not fitting any distinct population. In practical use scenarios, the Gaussian Process can be guided towards realistic rotational profiles through physically motivated priors for the metric parameter $M$.


Asteroid 299 Thora has a sparse data set, 49 observations over 153 days, and is characterized as having a rotational period of 269.79 hour with a 68\% credible interval ranging from 269.00300003 to 270.8985369 hours. The 273-hour rotational period reported by \citet{299thora} falls within three standard deviations of the peak and upon visual inspection, ZTF observations folded at 273 hours produce a more tightly-constrained rotational profile with less scatter. This could imply that 273 hours is a more accurate rotational period for Asteroid 299, and that the current selection of observations don't lend themselves to this interpretation. Another possible explanation could be that the phase-angle was not fully normalized with OpenOrb. ZTF observations for Asteroid 299 show a slight average increase in magnitude over time (about a 0.1 magnitude rise over 153 days) which could be indicative of an inaccurate phase-angle model. Our model does not yet account for any changes in the average light curve magnitude. That aspect of the light curve is modeled with the mean function, which at the moment we assume is constant (see Section \ref{sec:future} for further discussion). Thus, if our phase-angle correction was calculated incorrectly, then, our assumption that the mean function is constant is false, and it is possible that the metric $M$ of the squared exponential covariance function is compensating for our error by modeling the biases in the light curve introduced by an incorrect phase-angle correction. In this case, the metric and amplitude of the squared exponential covariance function should be interpreted with caution.

\subsection{Limitations}\label{sec:limitations}

We note some of the inherent limitations to the Gaussian Process model. There are occasions where the model fails to characterize an asteroid, either because (1) large outliers are present in the data, (2) uncertainties associated with the magnitude measurements are large, or (3) the span of observations is too short. 

Outliers in data sets, several standard deviations removed from the mean, can cause Gaussian Processes to fail. In these cases, we often see marginalized parameter posteriors that reproduce their prior distributions, spanning several orders of magnitude or their entire range as specified by their prior (e.g. the period parameter cannot exceed one year, or 8000 hours, set by its prior). If we remove these outliers, the model is able to successfully characterize the light curve.

The Gaussian Process model \com{might fail} to generate a good model for light curves if an asteroid's magnitude uncertainty constitutes more than 10\% of the intrinsic magnitude variability. 821 Fanny is a prime example of this limitation. It has a 25\% relative magnitude uncertainty when compared to its intrinsic variability, suggesting the observations are not constraining the Gaussian Process model, allowing it to fit almost any period to the data. It is possible that the current viewing angle of 821 Fanny is such that the light curve amplitude is minimized, maximizing the relative magnitude uncertainty. There exists a degeneracy between light curve amplitude and viewing angle, where observing an asteroid pole-on will result in a very small light curve amplitude, whereas observing an asteroid equatorially will allow for the full light curve amplitude to be observed, in the case of an oblate tri-axial ellipsoid. Knowing an asteroid's pole orientation will reveal when an asteroid is being observed equatorially versus pole-on \citep{Vokrouhlicky2017}. Regardless, the current ratio between 821 Fanny's magnitude uncertainty and the intrinsic magnitude variability does not constrain the model enough to provide valuable results.

The model also does not perform well when given data that does not span multiple rotational periods. However, if the rotational period of an asteroid is undetermined, it is impossible to know if this criterion has been met. If the model suggests a rotational period close to the length of the observational window (typically within an hour), it is a strong indication that the observational window of the data is too small (e.g. nightly observations consistently produce rotational period estimates ~24 or 48 hours). These results should not be trusted, and we suggest adding a secondary source of data to increase the length of the observational window. In most cases, this occurs when data only spans a couple of hours from one night of observations. In this situation, the model is unable to find a suitable rotational period shorter than the length of the observations, indicating that no smaller period exists and that no other method would be able to find an appropriate period.

\subsection{Recommendations for Usage}

Gaussian Processes implemented with the covariance functions used here are generally computationally expensive and scale with the cube of the number of data points, $N^3$, since they require computation of both determinant and inverse of the $N$x$N$ covariance matrix.
This model should ideally be used in scenarios where faster methods like the LSP either fail to find a believable period or show multiple peaks with similar heights in the periodogram. In both situations, the Gaussian Process model adds probabilistic information and, as we've shown, might be able to better constrain the asteroid's properties. These scenarios are most likely to occur with sparse data sets, which makes inference computationally feasible.


We envision an implementation in a two-step process, where initial screening of a large survey of asteroid light curves is initially analyzed using fast, traditional methods of period detection. An automatic process then generates candidates for further studies: either asteroids that are of particular interest, cases where the inferred period deviates from previous measurements for that asteroid, or cases where traditional methods produce ambiguous results. These sources may then be followed up with the approach laid out here to produce better inferences for periods, as well as other other parameters such as amplitudes and measures of the variation of the rotational profile that traditional methods cannot provide. 

\com{As generative models, Gaussian Processes can produce probabilistic forecasts of future behaviour.  This can be employed to make predictions for when potential follow-up observations should be optimally scheduled, for example in order to rule out certain modes in the posterior probability distribution and improve inference of the period or other parameters such as the amplitude.}

\subsection{Future Work}\label{sec:future}


While the Gaussian Process model constrains both the periodicity and the profile evolution as a result of the changing phase-angle, there is currently no term explicitly included in the Gaussian Process for modeling the overall phase-angle change of the light curve. Here, we normalized any variation in magnitudes resulting from the phase-angle out of the light curve. We accomplish this by modeling the expected ZTF-observed asteroid light curves using OpenOrb and then subtracting this model from the observed data. Future work would see an overall phase-angle term added to the model through a time-dependent mean function. This allows the model to correctly incorporate uncertainties in the phase angle model and help avoid situations where the period determination is strongly affected by biases introduced in an earlier phase angle correction step.

If we observe an asteroid in multiple bands, it is necessary to expand the model to simultaneously fit multiple light curves of the same period with differing profiles. Future work would see the addition of multiple bands to the model, similar to how the LSP from Gatspy can fit multiple bands at once. 

Computational improvements to the model may include a sampler more tailored towards multimodal posterior distributions, as well as a more computationally efficient covariance functions. Because the affine-invariant sampler implemented in \textit{emcee} uses a subset of chains to generate proposals for new parameters in other chains, it is generally not well-suited to the problem considered here. A future exploration and comparison of sampling algorithms and implementations may significantly improve both inferences and computational efficiency. The covariance functions chosen in this work have a computational scaling of $\mathcal{O}(N^3)$. While reasonable for sparse light curves, this scaling quickly makes the model prohibitively expensive when aiming to provide comparisons on densely sampled light curves. Other periodic or quasi-periodic covariance functions with more favourable computational complexity exist \citep[e.g.][]{celerite}, and may provide a similarly appropriate model for asteroid light curves.


In this paper, we design the prior probability for the period based on the distribution of known asteroid periods in the LCDB. While this distribution is potentially biased towards lower periods because of observational biases, we also show that it can nevertheless infer the period for long-period asteroids. 
In the future, implementing more realistic priors for both the period $P$ and metric $M$ parameters, potentially tailored towards certain asteroid populations of interest, should lead to improved inferences for these asteroids, and will allow better population studies.

Additionally, while we primarily focus on the inference of the rotational period in this paper, we can also explore the physical meaning of other parameters. In particular the length-scale parameter $\Gamma$ provides a proxy for the overall reflective properties of the asteroid which could be exploited to explore the composition and albedo of different populations.

\section{Conclusion}

In the era of wide-field surveys like the Rubin Observatory's Legacy Survey of Space and Time and the Zwicky Transient Facility, sparse photometric measurements will constitute an increasing percentage of asteroid observations, particularly for currently unknown asteroids that have yet to be detected. These surveys open the door for large-scale population studies, but require inferring physical properties from individual asteroids within these vast data sets. At the same time, follow-up observations to supplement existing data may be prohibitively expensive in many cases. 

In this paper, we set out to add to the toolbox of methods that can infer periods and other relevant asteroid properties from sparse light curves, and show that the Gaussian Process model can infer the correct period for both short-period and long-period asteroids.\footnote{\textbf{The source code for the Gaussian Process model is available as an open-source Python package on GitHub \url{https://github.com/dirac-institute/asterogap}, with documentation at \url{https://dirac-institute.github.io/asterogap/}. The version of the code that produced the results in this paper is archived on Zenodo \citep{zenodo}.
}} By implementing both a periodic sine-squared exponential covariance function and a squared exponential covariance function, we are able to model complex rotational profiles with only a few dozen observations. 
We test our Gaussian Process model on sparse simulated asteroid observations as well as ZTF observations. We find that our model produces reliable parameter estimates as long as the data do not include the concerns put forth in Section \ref{sec:limitations}. We hope that the addition of this new model will allow astronomers to characterize asteroids whose properties were previously unconstrained.

\section*{Acknowledgments}

The authors wish to thank Josef \v{D}urech for helpful discussion in accessing the DAMIT asteroid model database in generating the synthetic asteroid lightcurves used in this study as well as help with the DAMIT synthetic asteroid lightcurve code.

Based on observations obtained with the Samuel Oschin Telescope 48-inch and the 60-inch Telescope at the Palomar Observatory as part of the Zwicky Transient Facility project. ZTF is supported by the National Science Foundation under Grant No. AST-1440341 and a collaboration including Caltech, IPAC, the Weizmann Institute for Science, the Oskar Klein Center at Stockholm University, the University of Maryland, the University of Washington, Deutsches Elektronen-Synchrotron and Humboldt University, Los Alamos National Laboratories, the TANGO Consortium of Taiwan, the University of Wisconsin at Milwaukee, and Lawrence Berkeley National Laboratories. Operations are conducted by COO, IPAC, and UW.

D.H.  is supported by the Women In Science Excel (WISE) programme of the Netherlands Organisation for Scientific Research (NWO). D.H. acknowledges support from the DIRAC Institute in the Department of Astronomy at the University of Washington. The DIRAC Institute is supported through generous gifts from the Charles and Lisa Simonyi Fund for Arts and Sciences, and the Washington Research Foundation. This work was supported by a Data Science Environments project award from the Gordon and Betty Moore Foundation (Award \#2013-10-29) and the Alfred P. Sloan Foundation (Award \#3835) to the University of Washington eScience Institute and by the eScience Institute.

This research has made use of NASA's Astrophysics Data System Bibliographic Services.

\software{
\texttt{numpy} \citep{numpy},
\texttt{matplotlib} \citep{matplotlib},
\texttt{george} \citep{george},
\texttt{emcee} \citep{emcee},
\texttt{scipy} \citep{scipy},
\texttt{pandas} \citep{pandas},
\texttt{h5py} \citep{h5py},
\texttt{corner} \citep{corner}
}
\bibliography{references}

\section{\com{Appendix}} \label{sec:appendix}

\end{document}